\documentclass{aastex631}

\usepackage{placeins}
\usepackage{xspace}
\usepackage[varg]{txfonts}
\usepackage{natbib}
\bibpunct{(}{)}{;}{a}{}{,}
\usepackage{graphicx}
\usepackage{xcolor}

\newcommand{\kms}{\ensuremath{\mathrm{km\ s^{-1}}}\xspace}
\newcommand{\NH}{\ensuremath{N_{\mathrm{H}}}\xspace}
\newcommand{\zv}{\ensuremath{z_{\mathrm{v}}}\xspace}

\newcommand{\slab}{{\tt slab}\xspace}

\newcommand{\amol}{{\tt amol}\xspace}
\newcommand{\hot}{{\tt hot}\xspace}
\newcommand{\pow}{{\tt pow}\xspace}
\newcommand{\bb}{{\tt bb}\xspace}

\newcommand{\combinespectra}{{\tt combine\_grating\_spectra}\xspace}
\newcommand{\sectors}{{\tt sectors}\xspace}
\newcommand{\xmm}{{XMM-\textit{Newton}}\xspace}
\newcommand{\chandra}{\textit{Chandra}\xspace}
\newcommand{\spex}{{\tt SPEX}\xspace}
\newcommand{\cloudy}{{\tt Cloudy}\xspace}
\newcommand{\xstar}{{\tt XSTAR}\xspace}

\newcommand{\HSTCOS}{{HST/COS}\xspace}

\newcommand{\emcee}{\textit{emcee}\xspace}

\newcommand{\fei}{\ion{Fe}{1}\xspace}
\newcommand{\feii}{\ion{Fe}{2}\xspace}
\newcommand{\feiii}{\ion{Fe}{3}\xspace}
\newcommand{\feiv}{\ion{Fe}{4}\xspace}
\newcommand{\si}{\ion{S}{1}\xspace}
\newcommand{\sii}{\ion{S}{2}\xspace}
\newcommand{\siii}{\ion{S}{3}\xspace}
\newcommand{\cii}{\ion{C}{2}\xspace}
\newcommand{\ciii}{\ion{C}{3}\xspace}
\newcommand{\ci}{\ion{C}{1}\xspace}
\newcommand{\oi}{\ion{O}{1}\xspace}
\newcommand{\oii}{\ion{O}{2}\xspace}
\newcommand{\oiii}{\ion{O}{3}\xspace}
\newcommand{\oiv}{\ion{O}{4}\xspace}

\newcommand{\nei}{\ion{Ne}{1}\xspace}
\newcommand{\neii}{\ion{Ne}{2}\xspace}
\newcommand{\neiii}{\ion{Ne}{3}\xspace}
\newcommand{\hi}{\ion{H}{1}\xspace}

\begin{document}

\title{Elemental abundances in the diffuse ISM from joint FUV and X-ray spectroscopy: iron, oxygen, carbon and sulfur }

\author[0000-0002-1049-3182]{I. Psaradaki}
\affiliation{MIT Kavli Institute for Astrophysics and Space Research, 70 Vassar Street, Cambridge, MA 02139, USA}
\affiliation{LSA, University of Michigan, 1085 S University Ave, Ann Arbor, MI 48109, USA}
\affiliation{SRON Netherlands Institute for Space Research, Niels Bohrweg 4, 2333 CA Leiden, the Netherlands}
\author{L. Corrales}
\affiliation{LSA, University of Michigan, 1085 S University Ave, Ann Arbor, MI 48109, USA}
\author{J. Werk}
\affiliation{University of Washington, Seattle, WA 98195, USA}
\author{A. G. Jensen}
\affiliation{University of Nebraska at Kearney Department of Physics and Astronomy 2502 19th Ave Kearney, NE 68849, USA}
\author{E. Costantini}
\affiliation{SRON Netherlands Institute for Space Research, Niels Bohrweg 4, 2333 CA Leiden, the Netherlands}
\author{M. Mehdipour}
\affiliation{Space Telescope Science Institute, 3700 San Martin Dr, Baltimore, MD 21218, USA}
\author{R. Cilley}
\affiliation{LSA, University of Michigan, 1085 S University Ave, Ann Arbor, MI 48109, USA}
\author{N. Schulz}
\affiliation{MIT Kavli Institute for Astrophysics and Space Research, 70 Vassar Street, Cambridge, MA 02139, USA}
\author{J. Kaastra}
\affiliation{SRON Netherlands Institute for Space Research, Niels Bohrweg 4, 2333 CA Leiden, the Netherlands}
\author{J. A. Garc\'ia}
\affiliation{X-ray Astrophysics Laboratory, NASA Goddard Space Flight Center, Greenbelt, MD 20771, USA}
\affiliation{Cahill Center for Astrophysics, California Institute of Technology, Pasadena, CA 91125, USA}
\author{L. Valencic}
\affiliation{Johns Hopkins University, 3400 N. Charles St., Baltimore, MD 21218, USA}
\author{T. Kallman}
\affiliation{X-ray Astrophysics Laboratory, NASA Goddard Space Flight Center, Greenbelt, MD 20771, USA}
\author{F. Paerels}
\affiliation{Columbia Astrophysics Laboratory and Department of Astronomy, Columbia University, 550 West 120th St., New York, NY 10027, USA}

\begin{abstract}

In this study, we investigate interstellar absorption lines along the line of sight toward the galactic low-mass X-ray binary Cygnus X-2. We combine absorption line data obtained from high-resolution X-ray spectra collected with \chandra and \xmm satellites, along with Far-UV absorption lines observed by the Hubble Space Telescope's (HST) Cosmic Origins Spectrograph (COS) Instrument. Our primary objective is to understand the abundance and depletion of oxygen, iron, sulfur, and carbon. To achieve this, we have developed an analysis pipeline that simultaneously fits both the UV and X-ray datasets. This novel approach takes into account the line spread function (LSF) of \HSTCOS, enhancing the precision of our results.
We examine the absorption lines of \feii, \sii, \cii, and \ci present in the FUV spectrum of Cygnus X-2, revealing the presence of at least two distinct absorbers characterized by different velocities. Additionally, we employ \cloudy simulations to compare our findings concerning the ionic ratios for the studied elements. We find that gaseous iron and sulfur exist in their singly ionized forms, Fe II and S II, respectively, while the abundances of CII and CI do not agree with the Cloudy simulations of the neutral ISM.
Finally, we explore discrepancies in the X-ray atomic data of iron and discuss their impact on the overall abundance and depletion of iron.

\end{abstract}

\section{Introduction} 

The interstellar medium (ISM) is an important component of our Galaxy --- it contributes to many astrophysical processes and for the formation of new stars. The ISM evolves dynamically and is divided into different phases. The \textit{neutral phase} contains atomic gas at temperatures from $\rm \sim 10^{2} \ K$ to $\rm \sim 10^{3.7} \ K$, and \textit{molecular gas} can be found either in gravitationally bound clouds or in the diffuse ISM. These are typically cool regions ($\rm \sim 10 \ K$) and their density can vary from $\rm\sim 1000 \ cm^{-3}$ up to $\rm\sim 10^{6} \ cm^{-3}$. Finally, the \textit{ionized phase} describes the ISM regions with temperatures from $\rm \sim 10^{4} \ K$ up to $\rm \sim 10^{5.5} \ K$, which is called warm ionized medium (WIM) or hot ionized medium (HIM) respectively \cite[e.g.][]{Drainebook, Tielens2001}. 
A multi-wavelength approach provides the means to better understand the structure of the interstellar medium. UV absorption spectroscopic observations can probe gas-phase abundances through resonance transitions, while X-rays can also provide spectroscopic information about the solid phase. In this work, we use the combination of UV and X-ray spectra to better determine the elemental abundances in the ISM. \\

\subsection{Oxygen}
Oxygen is the most abundant cosmic element after H and He and the amount depleted into dust grains is highly variable with ISM phase. The overall estimate of the oxygen budget in the interstellar medium (ISM) remains highly uncertain, as noted by \cite{Jenkins2009}. Although it is anticipated that oxygen may experience some depletion into dust, a considerable portion of it appears to be absent from the gaseous phase, without a comprehensive explanation. The combined contribution of carbon monoxide (CO), ices, silicate and oxide dust particles are insufficient to fully account for the missing oxygen in the denser regions of the ISM, particularly at the interface where the diffuse and dense ISM meet, as highlighted by \cite{Whittet2010} and \cite{Poteet}. Oxygen has been extensively studied in the literature using high-resolution X-ray spectroscopy of the O K-edge (\citealt{Takei2002}, \citealt{Juett2004}, \citealt{deVries2009}, \citealt{Pinto2010, Pinto2013}, \citealt{Costantini2012}, \citealt{Gatuzz2014, Gatuzz2016}, \citealt{Joachimi2016}, \citealt{Eckersall2017}). In \cite{Psaradaki2020, Psaradaki2022} we studied the oxygen abundance in both gas and solids through the O K-edge. We found that about 10–20\% of the neutral oxygen is depleted into dust, and that in the diffuse sightlines the oxygen abundance is consistent with or slightly above the Solar value. In this work, we examine the X-ray and UV features simultaneously.

\subsection{Iron}
Iron is a major constituent in most dust grain models, as more than $90\%$ of the total iron is depleted from the gas-phase; the remainder is presumably locked up in dust grains (e.g. \citealt{Savage1979}, \citealt{Jenkins1986}, \citealt{Snow2002}, \citealt{Jenkins2009}). More than $\rm 65 \% $ of the iron is injected in the ISM in the gaseous form by Type Ia supernovae, therefore most of the Fe dust growth is expected to take place in the ISM (\citealt{Dwek2016}). However, the exact composition of Fe-bearing grains as well as the exact amount and form that iron takes in the ISM is still unclear. 
Iron is expected to be present in silicate dust grains but it could also exist in pure metallic nanoparticles (e.g. \citealt{Kemper2002}) or even as metallic inclusions in glass with embedded metal and sulphides (GEMS), suspected to be of interstellar origin (e.g \citealt{Altobeli}, \citealt{Bradley1994}, \citealt{Ishii2018}). The possibility of iron sulfides is discussed in more detail below.

UV and optical observations find that the remaining gas-phase Fe is primarily in the form of \feii in neutral regions of the ISM (\citealt{Snow2002}, \citealt{Miller2007}, \citealt{Jensen2007}). This is because the ionisation potential of \fei is 7.87 eV and it can be ionised by photons coming from the interstellar radiation field (ISFR), with energies between 7.87 eV and the Lyman limit at 13.6 eV. The ionisation potential of \feii is 16.18 eV, which lies above the cut-off energy of the ISFR and thereby unlikely to get ionised. In $\rm H_{II}$ regions, the gas-phase iron should be as a mix of \feii, \feiii and \feiv. In $\rm H_{II}$ regions, some \fei may exist, but due to depletion rates as high as 99\% found in cold neutral regions \cite[e.g.][]{Savage1996}, it is more likely to be in dust grains.  

\subsection{Sulfur}
The degree to which interstellar sulfur is depleted is still a matter of debate (\citealt{Jenkins2009}). In the diffuse ISM sulfur is expected to have modest depletion (\citealt{Costantini2019} and references therein). However, in denser regions, such as molecular clouds, sulfur can be included in aggregates such as H$_2$S and SO$_2$ (\citealt{Duley1980}). Sulfur in dust has been detected near C-rich AGB stars, planetary nebula (\citealt{Hony2002}) and protoplanetary disks (\citealt{Keller}). Solid Fe-S compounds are abundant in planetary system bodies, such as interplanetary dust particles, meteorites and comets \cite[e.g.][]{Wooden2008}. 
The presence of sulfur in dust grains can also be associated with GEMS (\citealt{Bradley1994}), where the FeS particles are concentrated on the surface of the glassy silicate. Metallic Fe particles embedded in a silicate matrix have become a popular model for explaining the depletion patterns of the ISM (e.g. \citealt{Zhukovska2018}).
The Stardust mission revealed sulfur in the form of FeS, suspected to be of ISM origin (\citealt{Westphal2014}). This evidence revitalizes the idea that sulfur could be present in dust species, also in less dense ISM environments (\citealt{Costantini2019}). However, it has been shown that GEMS are a less favored candidate of interstellar dust (\citealt{Keller}, \citealt{Keller2011}, \citealt{Westphal}).

In recent X-ray studies of interstellar Fe absorption, appreciable quantities of iron-sulphide material like troilite (FeS), pyrite Peru (FeS$_2$), and ferrous sulfate (FeSO$_4$) are not found \citep[][Corrales et al., submitted]{Psaradaki2022}. Moreover, \citet{Gatuzz2024} carried out a recent X-ray study of the sulfur K-edge. The authors estimated column densities of ionic species of sulfur along with column densities of dust compounds for a sample of 36 low-mass X-ray binaries. Upper limits were obtained for most sources including the dust components. However, they found that the cold-warm column densities tend to decrease with the Galactic latitude, while no correlation with distances or Galactic longitude. \si has an ionization potential of 10.36 eV, below the Lyman-limit, so the majority of gas-phase S in the neutral medium is expected to be in the form of \sii. This work examines \sii gas through the far-UV triplet transition at 1250.6 \AA, 1253.8 \AA \ and 1259.5 \AA. 

\subsection{Carbon}
Carbon is also suspected to be a major constituent of interstellar dust grains, however we have limited knowledge about the amount of carbon that is locked-up in dust grains (\citealt{Jenkins2009}). It has been suggested that carbon, constitutes around
20\% of the total depleted mass in the Galaxy (\citealt{Whittet2003}, \citealt{Draine_2021}).
Its depletion covers a relatively narrow range of values, showing that it is not a strong function of environmental density (\citealt{Costantini2019}). The majority of carbon should be locked in graphite grains, providing a likely explanation
for the 2175 \AA \ emission feature (\citealt{Draine1989}, \citealt{Draine2003}, and references therein). 
However, concerns have been raised regarding the insufficiency of carbon depletion to account for the observed optical properties of interstellar dust (\citealt{Kim1996}, \citealt{Mathis1998}, \citealt{Dwek1997}).
A broad interstellar absorption feature at 2175\AA \ as well as narrow band emission features in the FIR are attributed to polycyclic aromatic hydrocarbons \cite[e.g.][]{Draine1989, Draine2003}. Various carbonaceous grain compositions proposed include graphite, hydrogenated amorphous carbon, and silicates with a carbonaceous mantles \citep{Duley1989, Weingartner2001, Zubko2004, Jones2017, Costantini2019}. Carbon could be also locked in nano-diamonds, which could be created from graphite and amorphous carbon grains in high pressure ISM environments, for example around shocks (\citealt{Tielens1987}). Nano-diamonds are also found in meteoritic material, and isotopic ratios imply that they are not from the Solar System. However, our knowledge of the actual depletion of carbon -- and thereby the total amount of carbonaceous interstellar dust -- is still an enigma. 

Carbon spectroscopy of Galactic sources is usually challenging in the X-rays due to very high absorption as well as the relative insensitivity of modern X-ray instruments near the C K photoelectric absorption edge at 0.3~keV. However, it is possible for very low column density sight lines. \citealt{Gatuzz2018} studied the C K-edge using high-resolution \chandra spectra of four novae during their super-soft-source phase. The authors detected resonances of \cii $\rm K\alpha$ as well as the \ciii $\rm K\alpha$ and $\rm K\beta$ transitions. 
Moreover, simultaneous examination of the X-ray and UV spectrum of the extragalactic source Mrk 509 (\citealt{Pinto_UV}) suggests that most of the neutral carbon is locked-up in dust, while the bulk of \cii comes from the warm-photoionised phase. 
In this work, we study gas-phase carbon in the far-UV through the \ci and \cii transition at 1328.8 \AA \ and 1335.7 \AA \ respectively. \\

We use the joined information from X-ray data, through \chandra and \xmm satellites, and Far-Ultraviolet (FUV) data from the Cosmic Origins Spectrograph (COS) on board the Hubble Space Telescope (HST) in order to understand the abundance and depletion of oxygen, iron, sulfur and carbon. In the last decades, high-resolution X-ray absorption spectroscopy has proven to be a powerful tool for studying the interstellar medium (e.g. \citealt{Wilms, Takei2002, Ueda, Juett2004, Garcia2011, Pinto2010, Pinto2013, Costantini2012, Schulz, Gatuzz2016, Joachimi2016, Yang2022}). In particular, in the X-rays we are able to study the solid phase composition of highly depleted elements, such as iron in the line of sight towards a bright background source.

X-ray absorption fine structures (XAFS) are spectroscopic features observed near the photoelectric absorption edges of solid material (dust), and their shape is the ultimate footprint of the chemical composition, size, and crystallinity \cite[e.g.][]{Newville, Lee2005, Zeegers2017, Zeegers2019, Rogantini2018, Rogantini2019, Corrales2016, Corrales2019, Costantini2022, Psaradaki2020, Psaradaki2022}. 
However, the abundance and absorption strength of \feii, likely the largest repository of gas-phase iron, is difficult to constrain from the X-ray band alone \citep[][Corrales et al., submitted]{Psaradaki2022}. 
In this pilot study, we use the joint information of the FUV and X-ray absorption spectra of the low mass X-ray binary (LMXB) Cygnus X-2 to study both the gas and dust component of the ISM, providing the most comprehensive means possible to determine the abundances and depletion of prevalent interstellar elements.
Cygnus X-2 is a bright X-ray source with a moderate column density ($\rm 2\times 10^{21}\ cm^{-2}$) and high flux ($\rm 2.3 \times 10^{-9} erg \cdot s^{-1} \cdot cm^{-2}$ in the 0.3–2 keV band), making an excellent target to study the diffuse interstellar medium in the O K and Fe L-edges. This sight-line also exhibits a rich FUV spectrum with absorption signatures from the ISM. The distance of the source has been estimated to be around 7–12 kpc (\citealt{Cowley}, \citealt{McClintock}, \citealt{Smale}, \citealt{Yao2009}).
This paper is organised as follows. In Section \ref{reduction} we present the \HSTCOS, \chandra and \xmm data used in this study and their reduction processes. In Section \ref{uv_fit} we describe the adopted method for analysing the FUV spectra, and in Section \ref{xray_fit} and \ref{oxygen} the spectral fitting to the X-ray data. Finally in Section \ref{discussion} we discuss the results, and give our conclusions in Section \ref{conclusions}.

\section{Data reduction}
\label{reduction}

The \HSTCOS (\citealt{COS_handbook}) datasets for Cygnus~X-2, described in Table~\ref{observation_log}, were downloaded from the MAST Portal  archive\footnote{\url{https://mast.stsci.edu/portal/Mashup/Clients/Mast/Portal.html}}. All the data-sets were obtained using the G130M filter. They consist of separate files for each of the two FUV detector segments, segment A and segment B. We reduce and combine the data-sets for both segments using the documented instructions in the Hubble Space Telescope User page\footnote{\url{https://hst-docs.stsci.edu/cosdhb/chapter-5-cos-data-analysis/5-1-data-reduction-and-analysis-applications}}.

We further obtained the Cygnus~X-2 datasets from the \xmm\ Reflection Grating Spectrometers (RGS, \citealt{denHerder}), which has a resolving power of $R=\frac{\lambda}{\Delta \lambda } \gtrsim 400$ and an effective area of approximately $\rm 45 \ cm^{2}$ in the spectral region of interest.  The datasets were downloaded from the \xmm\ archive\footnote{\url{http://nxsa.esac.esa.int/nxsa-web/}} (Table~\ref{observation_log}) and reduced using standard calibration procedures of the Science Analysis Software, $\textit{SAS}$ (ver.18). We created the event lists by running the $rgsproc$ command. Then, we filtered the RGS event lists for flaring particle background using the default value of 0.2 counts/sec threshold. We excluded the bad pixels using \texttt{keepcool=no} in the SAS task $rgsproc$. Moreover, when the spectral shape does not vary through different epochs and the spectra can be superimposed, we combined the data using the SAS command $\textit{rgscombine}$. This allowed us to obtain a single spectrum with higher signal-to-noise ratio. 

The \chandra \ observations used in this work were downloaded from the Transmission Grating Catalogue\footnote{\url{http://tgcat.mit.edu/tgSearch.php?t=N}} (TGCat, \citealt{tgcat}). \chandra\ carries two high spectral resolution instruments, the High Energy Transmission Grating \cite[HETGS,][]{hetgs} and the Low Energy Transmission Grating \cite[LETGS,][]{letgs}. The HETGS consists of two sets of gratings, the Medium Energy Grating (MEG) and the High Energy Gratings (HEG). In this study we are mainly interested in Fe L shell photoelectric absorption edges. We therefore used HETGS/MEG due to its high throughput and spectral resolution ($R=\frac{\lambda}{\Delta \lambda } \gtrsim 660$) around the Fe L edges. For each observation, we combine the positive and negative orders of dispersion using the X-ray data analysis software, CIAO (version 4.11, \citealt{ciao}). 
The persistent emission of the source is steady, and we therefore combine the different observations using the CIAO tool \combinespectra.

\begin{table}
\caption{Observation log. CC is an acronym for Continuous Clocking mode. } 
\centering
\label{observation_log}    
\begin{tabular}{c c c c }     
\hline\hline            
Satellite                       & obs ID         &     instrument/mode   &   exposure time (ks)   \\
\hline   
\hline   
{HST}     &   lb2m02010          &    COS/FUV/G130M                     &     7                \\
 							 &   lb2m03010         &    COS/FUV/G130M                     &      7            \\
 							 &   lb2m04010         &    COS/FUV/G130M                    &      7            \\

\hline                
{\xmm}           &   0303280101   &   RGS                 &    32            \\
 							    &   0561180501    &  RGS                   &  24                 \\
\hline                
{\chandra}     &     8170         &    HETGS/CC          &    65       \\
 							    &      8599      &    HETGS/CC         &      60             \\

\hline                
\end{tabular}
\end{table}

\section{The Far-UV spectrum}
\label{uv_fit}

The \HSTCOS instrument covers a wavelength range of the FUV absorption lines that is useful for this study. In particular we examine the COS spectrum of Cygnus X-2 in the range of 1132-1280 \AA \ and 1288-1430 \AA \ for segment B and A respectively. We studied in detail the absorption features of \feii \ (1142.36 \& 1143.22 \AA), \sii \ (1250.6 \& 1253.8 \AA), \ci \ (1328.7 \AA), \cii \ (1335.7 \AA) and \oi (1302.16 \AA).

\FloatBarrier
\begin{figure}
\centering
  \includegraphics[width=0.5\linewidth]{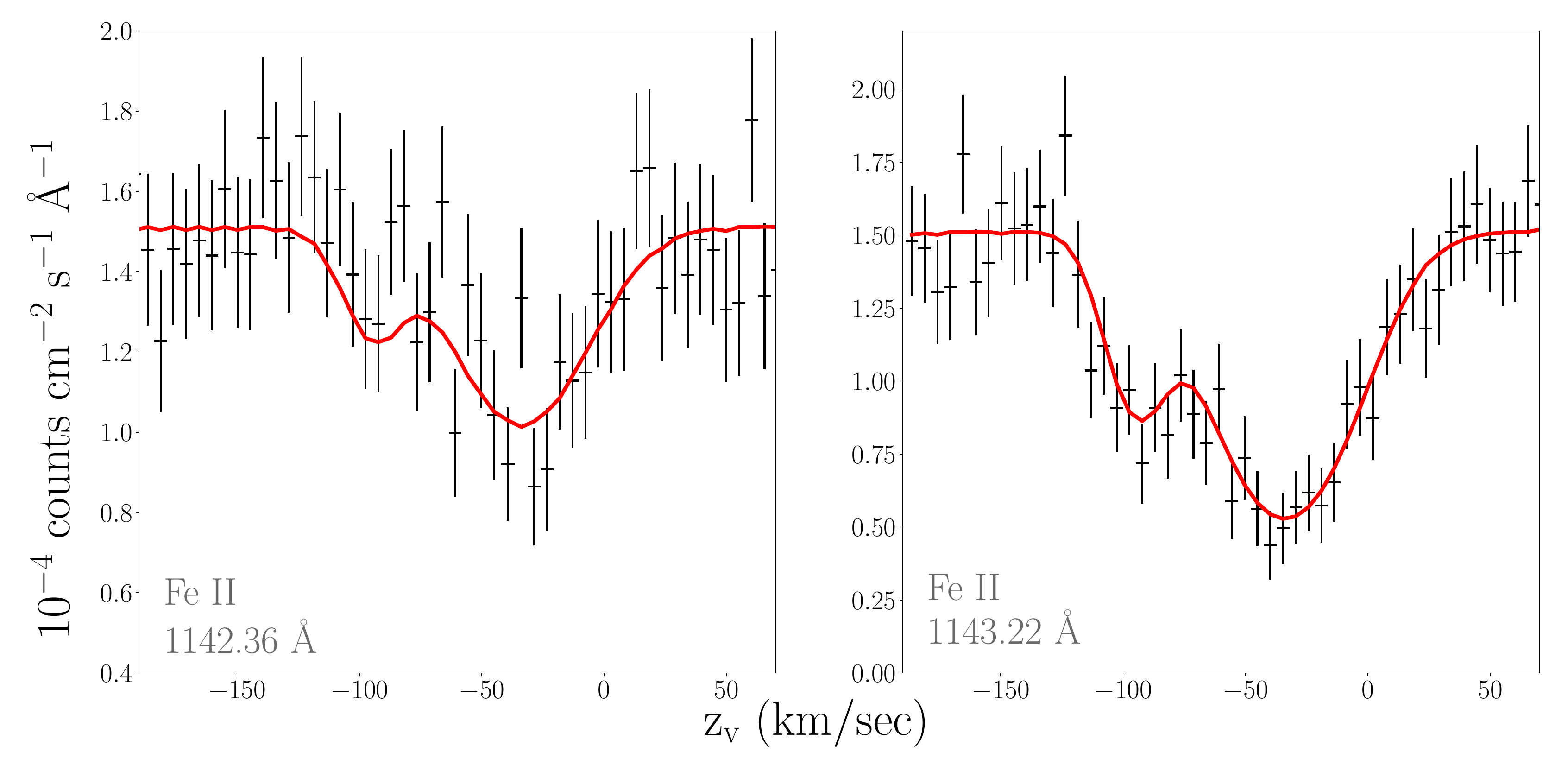}
  \includegraphics[width=0.5\linewidth]{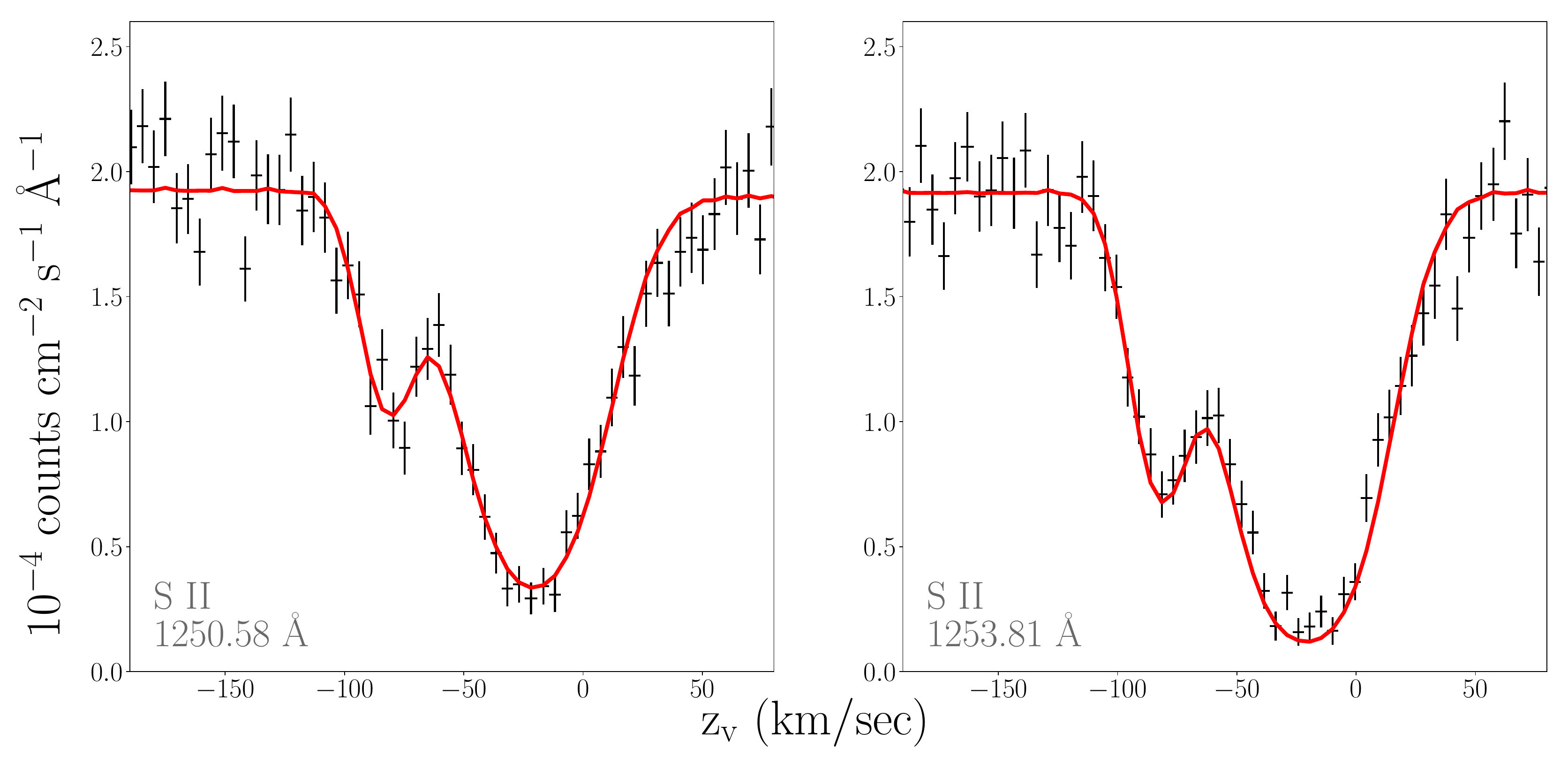}
  \includegraphics[width=0.5\linewidth]{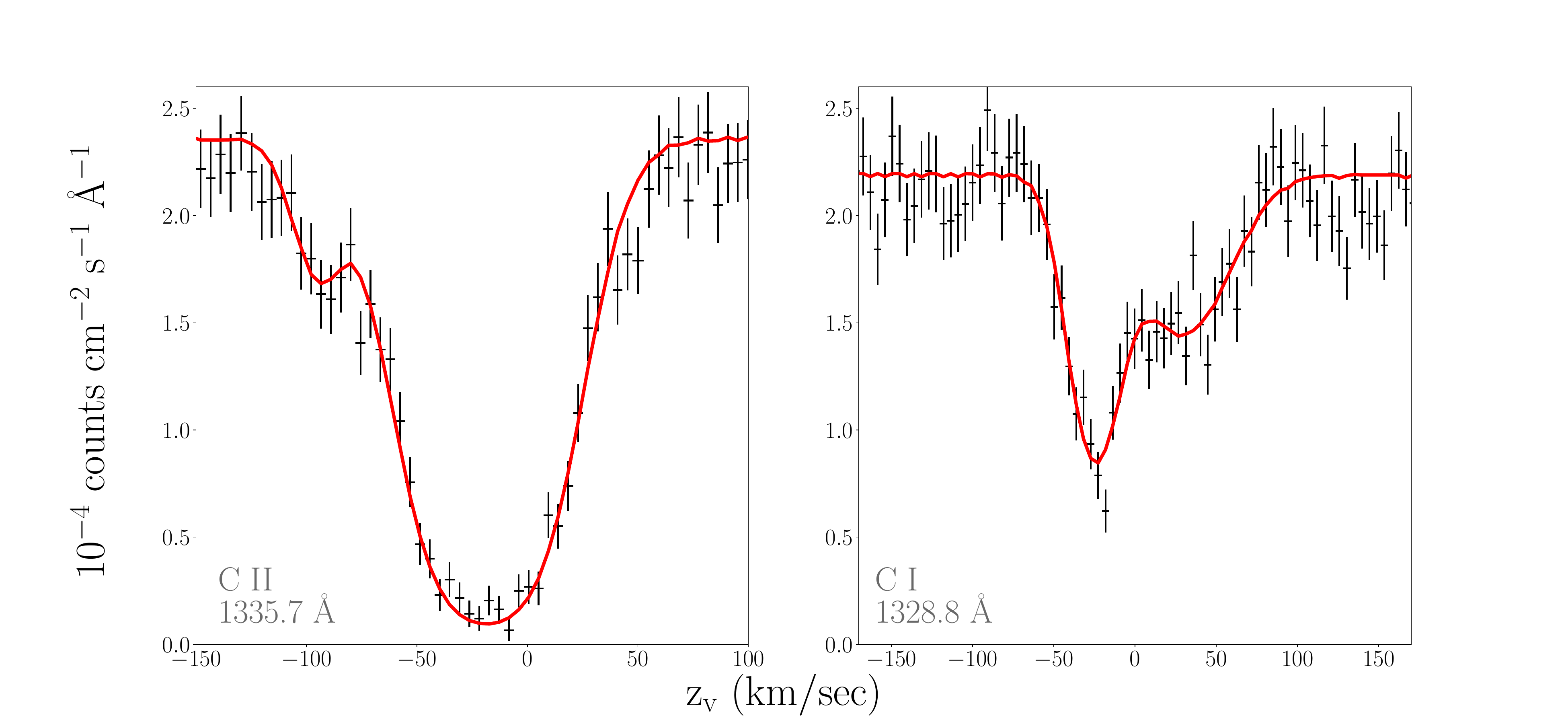}
\caption{COS/FUV absorption lines and best fit model for \feii, \sii, \ci and \cii in velocity space.}
 \label{fig:cosdata}
\end{figure}

\begin{table}
\caption{Oscillator strength ($\rm f_{ij}$) and rest frame wavelength ($\rm \lambda$) used in this study based on \citet{Morton1991}. }  
\label{tab:morton}    
\begin{center}
\begin{tabular}{c c c c }     
\hline\hline            
Ion                          &   $\rm \lambda$  (\AA)         &     $\rm f_{ij}$    \\
\hline   
\hline   
 {\feii}       &       1142.36      &   $5.00 \cdot 10^{-3} $     \\
 		    &     1143.22        &  $1.33 \cdot 10^{-2} $       \\
\hline                
{\sii}        &       1250.58      & $5.45 \cdot 10^{-3} $          \\
 		 &       1253.81      & $1.08 \cdot 10^{-2} $          \\
\hline                
{\ci}          &  1328.83             & $5.804 \cdot 10^{-2}$       \\
\hline                
{\cii}         &      1335.7          & $1.15 \cdot 10^{-1} $      \\
\hline
{\oi}         &      1302.16          & $4.88 \cdot 10^{-2} $      \\
\hline
\end{tabular}
\label{morton}
\end{center}
\end{table}

\begin{table}
\caption{Best fit parameters for all the observed ions and velocity components using \spex \ and \emcee.}  
\label{spexparams}    
\begin{center}
\begin{tabular}{c c c c }     
\hline\hline            
Ion                          &   $N_{i}$ \ ($\rm 10^{15}$ $\rm cm^{-2}$)         &     \zv \ (\kms)   & b (\kms)    \\
\hline   
\hline   
 {\feii}       &       $1.7\pm0.3$                  &   $-37\pm4$                &  $25\pm2$     \\
 		   &        $0.3\pm0.2$                        &     $-96\pm5$                   &   $10\pm1$        \\
\hline                
{\sii}         &          $7.4^{+1.3}_{-0.8}$              &    $-20\pm2$                &      $18\pm2$    \\
 		   &          $1.9^{+8}_{-0.7}$               &    $-84\pm2$                &    $5\pm3$       \\
\hline               
{\ci}          &         $0.21^{+0.09}_{-0.05}$          &     $-25\pm3$               &     $8^{+5}_{-1}$     \\
 		   &         $0.15^{+0.06}_{-0.04}$             &     $26^{+10}_{-12}$     &     $30^{+12}_{-15}$    \\
\hline                
{\cii}         &           $0.47\pm0.03$                     &      $-19\pm2$             &    $26\pm2$   \\
                &           $<0.04$          &      -95 (frozen)           &    $<21$    \\
\hline
\end{tabular}
\end{center}
\end{table}

We use the spectral models of pySPEX\footnote{\url{https://spex-xray.github.io/spex-help/pyspex.html}}, the python version of the software SPEctral X-ray and UV modelling and analysis, \spex \ (\citealt{Kaastra2018}), version 3.06.01. Even though \spex is primarily used for X-ray spectral analysis, its ability to model UV spectra with X-rays simultaneously has been demonstrated in the past (e.g. \citealt{Pinto2013}). We modified the \spex atomic databases to ensure that the rest-frame wavelengths and oscillator strengths for the ions of interest come from the same study (\citealt{Morton1991}). In Table \ref{morton} we present the wavelength and oscillator strength of each individual line used in this study. In the case of strong transitions and saturated lines, the derived column density was based on full Voigt profile fitting of the line features, which includes the damping wings. 
The exact value of the rest-frame velocity of each individual ion studied here, together with its oscillator strength, will be important for the calculation of the ionic column densities.   

\spex \ allows us to measure the column density of individual absorption lines using the \slab \ model. This model calculates the absorption by a slab of optically-thin gas, where the column densities of ions are fitted individually and are independent of each other. The free parameters are the ionic column density ($N_{i}$ in $\rm cm^{-2}$, where $i$ is the ion), the Doppler shift (\zv \ in \kms), and the rms broadening of the observed absorption line ($b$ in \kms). 
\spex is designed to work with X-ray calibration files, which are differently designed than those used to analyze UV data. Consequently, we developed a Python procedure that uses \spex to determine a physical model for the FUV dataset, then convolves that model with the COS line spread function (LSF) for the wavelength of interest following the \HSTCOS users' manual\footnote{\url{https://www.stsci.edu/hst/instrumentation/cos/performance/spectral-resolution}}. 
The LSF describes the light distribution at the focal plane as a function of wavelength in response to a monochromatic light source. The dominant effect in the observed spectrum is a broadening of the spectral features and the filling-in of saturated line features due to the finite resolution of the instrument. COS LSFs are known to have non-Gaussian wings and a model of the LSF is needed to perform accurate line profile fitting. 

We used the Monte Carlo Markov Chain (MCMC) analysis package \emcee \  (\citealt{emcee}) for the final spectral fitting. The \emcee\ package employs an ensemble sampler to probe the model parameter space.  It uses the Metropolis–Hastings algorithm which enables sampling from multi-dimensional distributions. 
We used the $\chi^{2}$ statistic (\citealt{Pearson1900}) to characterize the model likelihood for the UV datasets. We did not use priors, and for the burn-in phase we initiated the walkers using normal distributions around the initial values provided by an initial fit in \spex with a 25\% dispersion. We ran \emcee \ with 180 walkers and 65 steps for the burn-in phase, then ran the MCMC sampler for 260 steps to arrive at a posterior probability distribution. 
In the spectra obtained from \HSTCOS, we can observe distinct velocity components for all the lines under investigation (Figure \ref{fig:cosdata}). These lines reveal the presence of two absorbing clouds, each characterized by different velocity properties. Specifically, the absorption lines originating from the Fe II ion exhibit the presence of two blue-shifted components: one with a velocity shift of -37 ± 4 km/s and another with -96 ± 5 km/s. Similarly, the S II lines display two distinct components, one situated at -20 ± 2 km/s and the other at -84 ± 2 km/s relative to the rest-frame velocity. Furthermore, both \sii and \cii lines appear to be slightly saturated.

\begin{figure*}[ht]
\begin{centering}
\includegraphics[width=0.85\textwidth]{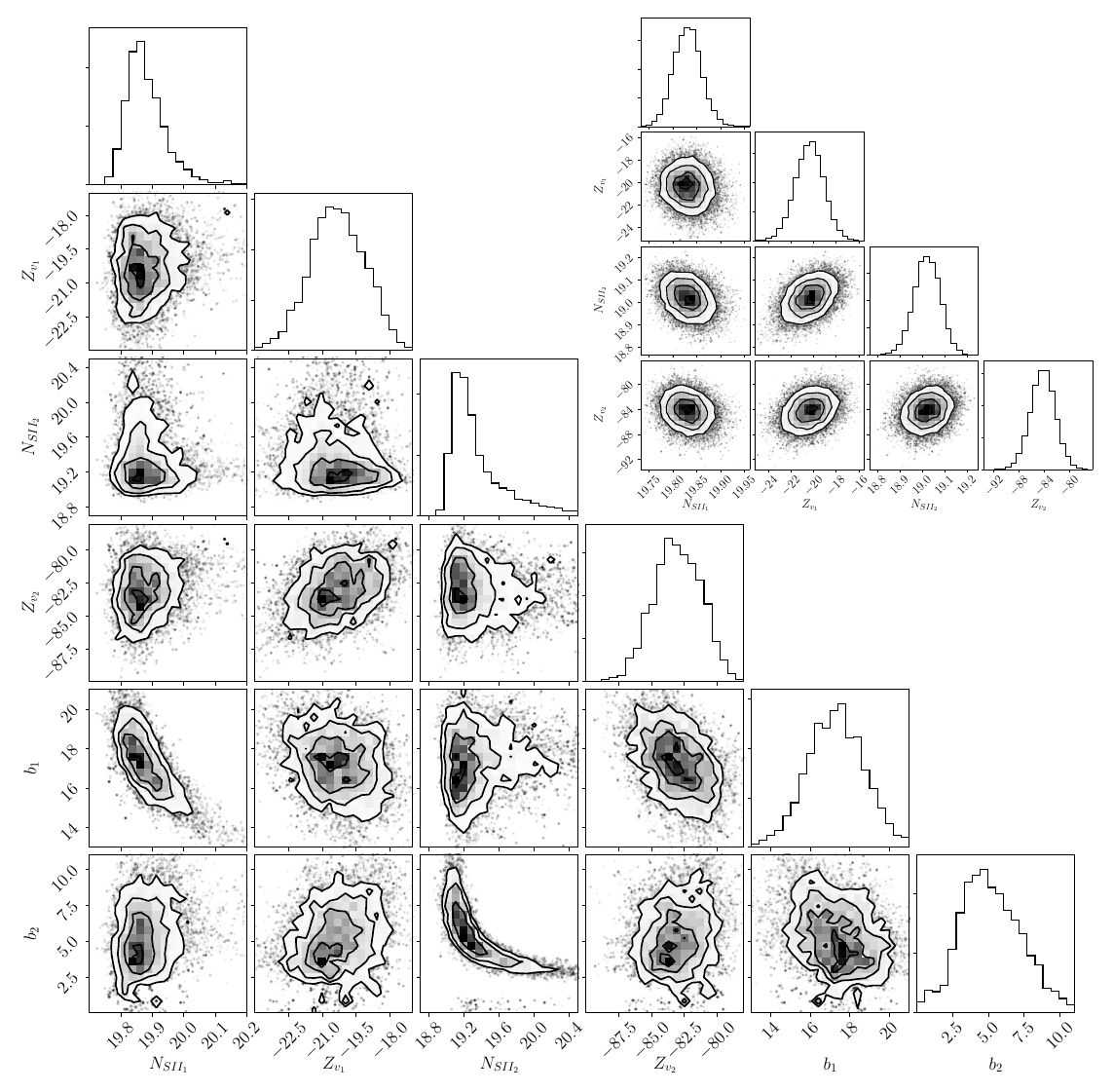}
\caption{The posterior distribution for \sii \ is plotted with two-dimensional histograms comparing each pair of free parameters. The contours represent the confidence level of $\rm 1\sigma$, $\rm 2\sigma$ etc. $\rm N_{SII}$ is the logarithmic ionic column density in $\rm cm^{-2}$, $\rm Z_{v}$ is the velocity shift of each component in km/sec, and $\rm b$ the velocity broadening in km/sec. The inset displays the identical posterior distribution, where the velocity broadening parameter is fixed to the fitted value.} 
\label{fig:SII_chains}
\end{centering}
\end{figure*}

Our first fitting procedure solved for the ionic column density of the two absorptions components ($N_{i}$), the line-of-sight velocity shift ($\rm z_{v}$) and their velocity broadening ($b$). Due to the fact that column density can be degenerate with the velocity broadening, we performed a second fit with the velocity broadening $b$ frozen to the best fitting value found in the first {\it emcee} run. This provides a stronger constraint on the ionic column densities. 
The best fit parameters for all ions with their 1$\sigma$ intervals are listed in Table \ref{spexparams}. The best fit to all FUV lines are presented in Figure \ref{fig:cosdata}. An example of the MCMC posterior distribution is presented in Figure~\ref{fig:SII_chains}, for the \sii \ lines. The illustrated corner plot shows the posterior distribution with two-dimensional histograms comparing each pair of free parameters including the velocity broadening, $b$. In the Figure~\ref{fig:SII_chains} inset, we present the posterior distribution for the same number of iterations and walkers, but this time $b$ is fixed to the fitted values. When the $b$ parameter is fixed, the posterior distribution exhibits a symmetrical distribution, whereas when the parameter is left free, we observe a degeneracy between the velocity broadening and the ionic column density.

\section{The X-ray spectrum: Fe L-edges}
\label{xray_fit}

The high-resolution X-ray spectrum of Cygnus X-2 includes the narrow absorption lines produced by neutral and ionised gas in the ISM around the photoabsorption edges of Ne K (13.5 \AA), O K (23 \AA), and Fe L (17.5 \AA). In this part of the study we are interested in the spectral region that contains the Fe L-edges, so we limit our fitting to the narrow range of 15-19 \AA. For this study, we combine the capabilities of both the \chandra and \xmm satellites.  

We use the available plasma models in \spex \ in order to fit the HETGS/MEG and RGS spectra of Cygnus X-2 jointly. We bin the data by a factor of 2 which improves the signal-to-noise while the data are still oversampling the spectral resolution of the instruments and we are not losing accuracy. We adopt $C$-statistics ($C_{\rm stat}$) to evaluate the goodness-of-fit (\citealt{Cash1979}, \citealt{Kaastra2017}). The models adopt the proto-solar abundance units of \cite{Lodders2009}. To take into account the continuum variability among the different data-sets, we use the $\sectors$ option in \spex. Each data-set can be allocated to a different sector, which allows us to fit the continuum parameters for each dataset independently. 
As we are fitting a relatively narrow energy band, the full shape of the spectral energy distribution cannot be constrained. Thus, we fit the continuum using a phenomenological power law ($\pow$ model in $\spex$) and a black body component ($\bb$). The free parameters consist of the slope and normalization of the $\pow$ component, and the temperature and normalisation of the $\bb$.

To take into account neutral Galactic absorption, we adopt the $\hot$ model of \spex \ (\citealt{dePlaa}, \citealt{Steenbrugge2005}). For a given temperature and set of abundances, this model calculates the ionisation balance and then determines all the ionic column densities scaled to the prescribed total hydrogen column density. At low temperatures ($\sim$ 0.001 eV $\sim$  10 Kelvin), the $\hot$ model mimics a neutral gas in collisional ionization equilibrium, and the free parameters are the hydrogen column density in the line of sight and the temperature ($kT$, where $k$ is the boltzmann constant). 
In the diffuse ISM, the gaseous phase iron is expected to be predominantly in the form of \feii (e.g. \citealt{Snow2002}, \citealt{Jensen2007}) due to ionization by the interstellar radiation field. To model \feii, we set the Fe abundance of the \hot neutral gas model to zero and replace it with the \slab \ model in \spex.  
The abundance of \feii is frozen to the value found from the fit of the \HSTCOS spectrum, shown in Table \ref{spexparams}. 

Around 90-99\% of interstellar Fe is known to be depleted into dust (e.g. \citealt{Dwek2016}, \citealt{Psaradaki2022}). We use the \amol \ model in \spex, which calculates the transmission of a dust component, and leave the dust column density as a free parameter. We use the recently implemented dust extinction cross sections for the Fe L-edges (\citealt{Psaradaki2021}, Costantini et al. in prep), computed from laboratory data and presented in \citet{Psaradaki2020, Psaradaki2021,Psaradaki2022}. The dust models have been computed using anomalous diffraction theory \cite[ADT,][]{van_de_Hulst} and assuming a Mathis-Rumpl-Nordsieck dust size distribution \cite[MRN,][]{mathis}. MRN follows a power-law distribution, $dn/da\propto a^{-3.5}$, where $a$ is the grain size with a minimum cut-off of $\rm 0.005 \ \mu m$ and a maximum cut-off of $\rm0.25 \ \mu m$. \citet{Psaradaki2022} found that amorphous pyroxene ($\rm Mg_{0.75}Fe_{0.25}SiO_{3}$) on average accounts for 80\% of the dust mass, with metallic iron taking up the remaining 20\%. This result on the silicate mixture of dust is broadly consistent with studies in infrared wavelengths (\citealt{Min2007}), and dust depletion studies (\citealt{Kostantopoulou2023}). 
In this context, it is important to clarify that we employ the term "amorphous" as a collective term for all non-crystalline materials. As explained in \citet{Psaradaki2020}, our amorphous samples exhibit a glassy nature, with their structure potentially retaining a short-range order of atoms. Nevertheless, the Si-K edge spectra of these amorphous samples exhibit a distinctively smooth profile, contrasting significantly with crystal samples (\citealt{Zeegers2019}).
We start by fitting the X-ray spectrum of Cygnus X-2 assuming this type of dust mineralogy. 

\begin{figure*} 
\centering
\includegraphics[width=0.45\textwidth]{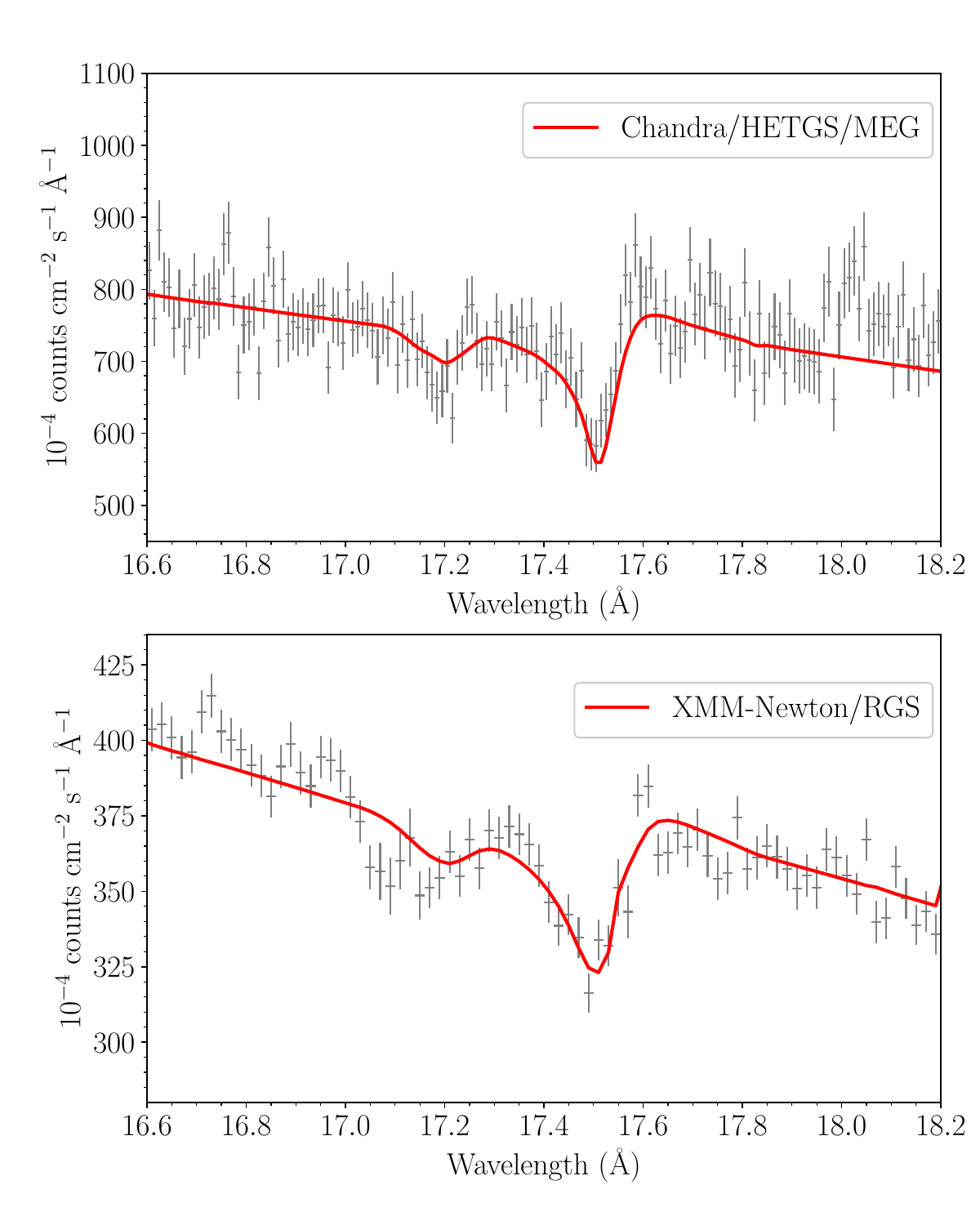}
\caption{Best fit in the Fe L-edges towards Cygnus X-2. Top: Chandra/HETGS spectrum. Bottom: \xmm/RGS spectrum.} 
\label{fig:iron_L}
\end{figure*}

The free parameters of our fit are the column density for each dust component, and the parameters of the continuum model including \NH.  The depletion of silicon and magnesium is fixed to be at least 90\% according to literature values (\citealt{Zeegers2019}, \citealt{Rogantini2019}) and the depletion of oxygen is constrained to be at least 20\% (\citealt{Psaradaki2022}). The best fit is shown in Figure \ref{fig:iron_L}, with $C_{stat}$/dof= 1098/736. As discussed in \cite{Psaradaki2022}, the remaining residuals around 17.6 \AA \ are possibly due to the MRN grain size distribution assumed in this study (Costantini et al. in prep). MRN provides a typical grain size distribution, while larger grain size is able to produce a larger scattering component of the extinction cross section \cite[e.g.][]{Corrales2016, Zeegers2017}. Dust size distributions larger than the MRN (larger maximum cut-off) will be examined in a follow-up paper. 

In summary, our analysis confirms that the dust column density is consistent with earlier findings (\citealt{Psaradaki2022}), demonstrating the reliability of our method. We find a dust column density of $\rm (7.9\pm1.6)\times 10^{16}\ cm^{-2}$. Moreover, we've integrated the \feii gas into our modelling, creating a consistent overall picture of how iron is distributed in the line of sight towards Cygnus X-2. We find that the majority of iron is in solids, while \feii accounts for the atomic phase iron, about 4\%. 

\section{Simultaneous fit of the X-ray and FUV spectrum: O K-edge}
\label{oxygen}

Neutral gas-phase oxygen is highly abundant in the ISM.  Large gas-phase column densities and the strong oscillator strength of the 1302 \AA \ absorption line of \oi found in the COS spectrum mean that this line is highly saturated. This typically puts the \oi absorption line in the transition region between flat and damped on a curve of growth (\citealt{Drainebook}). 
In the regime of the damped portion of the curve of growth, the core of the absorption line is totally saturated, but the “damping wings” of the line provide measurable partial transparency. This gives the possibility to measure the abundance of neutral oxygen through the COS spectrum,
and the only possibility for measuring its abundance comes from fitting the line profile, including the damping wings. 
Fortunately, gaseous \oi also has prominent absorption features in the X-ray, near the K shell photoelectric edge of oxygen, around 23 \AA \ (0.55 keV). These features are typically optically thin and closer to the linear portion of the curve-of-growth \citep{Juett2004}. Therefore we can more accurately constrain the gaseous abundance of \oi and solid phase components of interstellar oxygen through a simultaneous fit of the UV and X-ray datasets.
 
We limit our fit to the \oi bearing spectra from \HSTCOS and the X-ray portion of the O K-edge (19-25 \AA) using \xmm/RGS data. 
We used the \sectors option in \spex to fit with a different continuum model the two datasets, while the same ISM model is used in both sectors. For the FUV continuum, we use a phenomenological power law component (\pow), and for the X-ray continuum we adopted the best fit continuum parameters from \cite{Psaradaki2020}. We let the X-ray continuum free, but we kept the FUV continuum frozen to the initial values found from a preliminary fit in \spex.  
To model the \oi features in the X-rays, we employed the \spex\ \hot\ model with the plasma temperature frozen to the minimum value of 10 Kelvin. The \oi abundance is scaled from the model \NH column density, following proto-Solar oxygen values tabulated in \cite{Lodders2003}. \NH (and thereby the \oi column density) is left as a free parameter. We added a Gaussian prior on \NH of $\rm 2 \times 10^{21} \ cm^{-2} $, consistent with the work of \cite{Kalberla2005}, with a 10\% dispersion. We froze the depletion of Fe, Si and Mg to 0.1, according to the values found in \cite{Psaradaki2022, Zeegers2019, Rogantini2019}. The depletion of oxygen is a free parameter, but we add a Gaussian prior of 0.95 with 10\% dispersion to limit known degeneracies between this parameter and \NH. Finally, we let the velocity broadening parameter ($b$) and the velocity shift of the lines ($z_{v}$) to be free,  in order to determine the kinematics of the \oi line.

The O K-edge spectral region include transitions of other ions such as \oii and \oiii, along with highly ionized O that is likely intrinsic to the X-ray binary (e.g. \citealt{Juett2004}, \citealt{Pinto2010}, \citealt{Costantini2012}, \citealt{Gatuzz2016}, \citealt{Psaradaki2020}). These ions are included in the fit via the \slab model in \spex. The ionic column densities of \oii, \oiii, and \oiv lines are left as free parameters, while  the highly ionised lines are frozen to the values found in \cite{Psaradaki2020}. Oxygen absorption by dust is provided by the \amol model in \spex, which uses the dust extinction cross section computed from laboratory data described in \cite{Psaradaki2020}. We used the best fit compound from that work, amorphous pyroxene, $\rm (Mg,Fe)SiO_{3}$, as the only dust species in the fit, leaving the column density free. 

\begin{figure*} 
\centering
\includegraphics[width=0.42\textwidth]{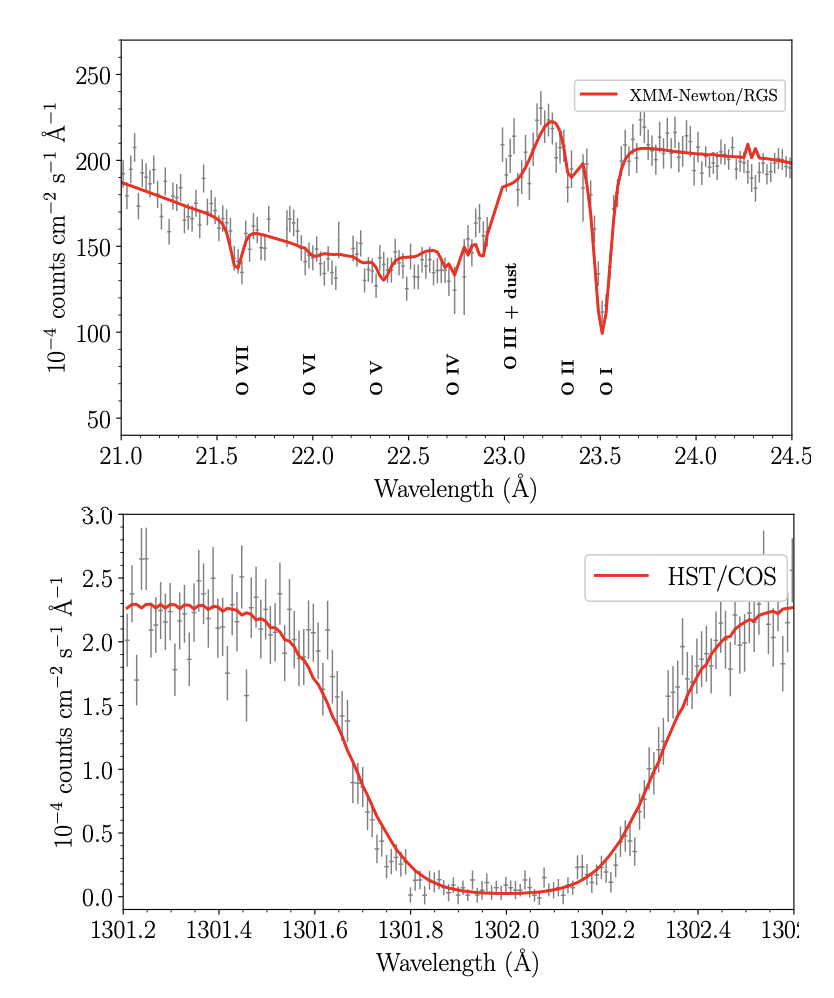}
\caption{Best fit in the O K-edge and \oi line profile in the \HSTCOS. Top: \xmm/RGS. Bottom: \HSTCOS} 
\label{fig:oxygen_sp}
\end{figure*}

\begin{figure}
 \label{fig:ox_posterior}

 \centering
  \includegraphics[width=1.0\linewidth]{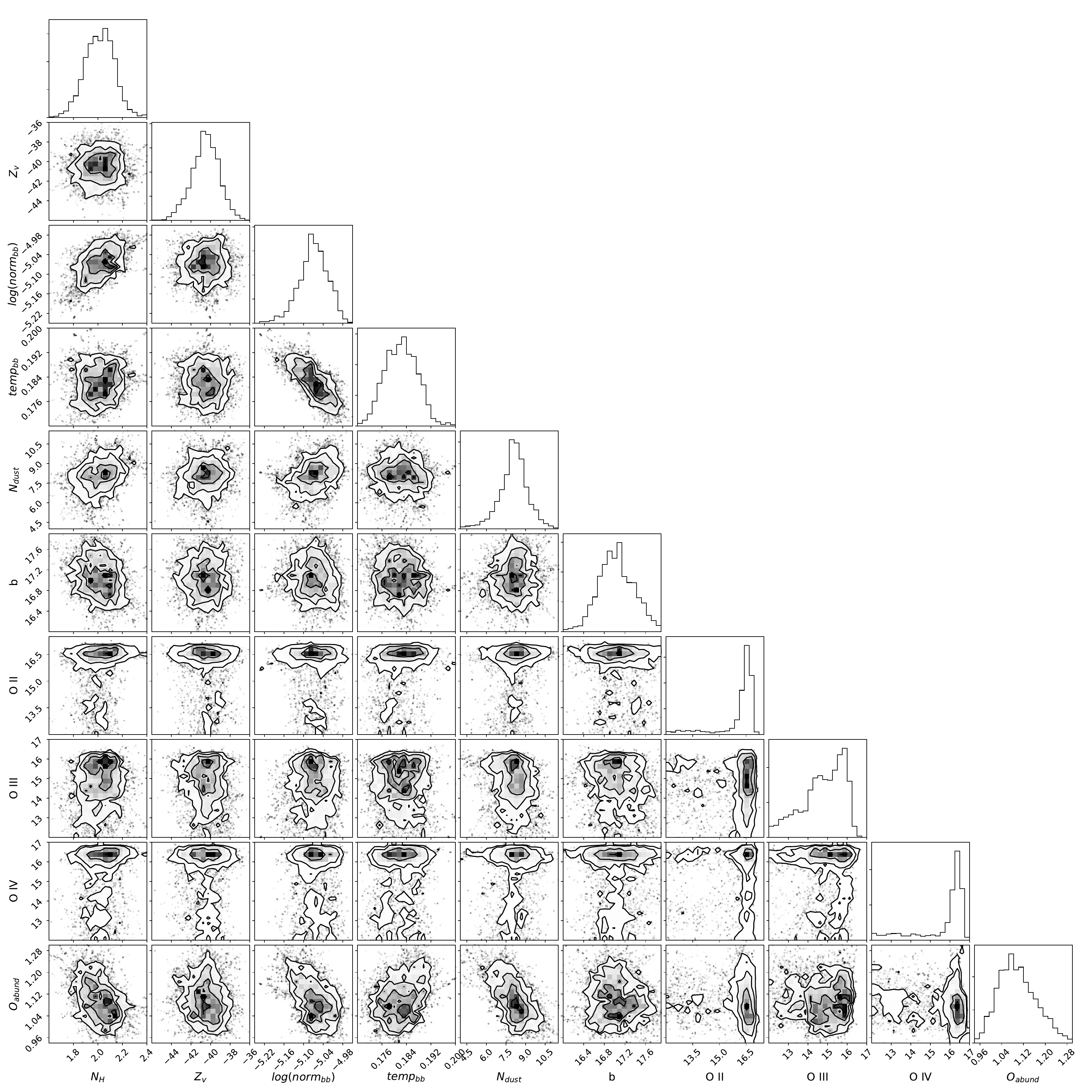}
\caption{Posterior distribution for the simultaneous FUV and X-ray spectrum in the oxygen region. The distribution compares each pair of free parameters of the fit with two-dimensional histograms. The parameters in the caption correspond to the following quantities: \NH is the neutral hydrogen column density in units of $\rm 10^{21}\ cm^{-2}$ along the line of sight, $\rm Z_{v}$ is the velocity shift in $\rm km \cdot s^{-1}$, $\rm log(norm_{bb})$ is the logarithm of the black body normalisation, $\rm N_{dust}$ is the total column density of dust in $\rm 10^{18}\ cm^{-2}$, $b$ refers to the line velocity broadening in $\rm km \cdot s^{-1}$, \oii, \oiii and \oiv correspond to the logarithmic ionic column densities in $\rm cm^{-2}$, and $O_{abund}$ is the O abundance scale factor relative to \cite{Lodders2009}.} 
\end{figure}

We combine our newly developed analysis pipelines, described in Sections~\ref{uv_fit} and \ref{xray_fit}, to fit the X-ray and FUV spectra simultaneously.  
The log-likelihood function used in this fitting procedure combines the chi-square statistic for the ultraviolet (UV) data with the Cash statistic obtained from the X-ray data. This method allows for a comprehensive and robust evaluation of the data from various spectral regions. To achieve convergence of the MCMC chain we run the code for 200 iterations and 80 walkers for the 100 free parameters of the study. For the burn-in phase we initiated the walkers in a normal distributions around the best fit parameters found via \spex and run MCMC for 50 steps. In Table \ref{tab:oxygen} we summarize all the free parameters in this study and the best fit values. We present the best fit in Figure \ref{fig:oxygen_sp} and the corresponding posterior distribution in Figure \ref{fig:ox_posterior}. The best fit suggests a slight over-abundance of oxygen, $1.1\pm0.1$, compared to the \cite{Lodders2009} abundance table. 

\begin{table}
    \centering
    \begin{tabular}{c c c }
\hline\hline
        component    &       parameter &   value                             \\
\hline
\hline
\bb         &    T (keV)                              & $ 0.18\pm0.01$        \\
         &        log($\rm norm_{bb}$) ($\rm 10^{16} \ m^{-2}$)  & $-5.07\pm0.04$   \\ 
\hline
  \hot   &       \NH  \ ($\rm 10^{21} \ cm^{-2}$)     &  $2.0\pm0.1$      \\
         &         $\rm z_{v} \ (km/sec)$             &  $-40\pm2$        \\
          &         b  (km/sec)                       &   $17.0 \pm0.4$    \\
         &        $O_{abund}$                         &   $1.1\pm0.1$     \\
         &        log(\oi)  ($\rm cm^{-2}$)    &      $18.04\pm0.03$   \\
\hline
\slab     &     log(\oii) \ ($\rm cm^{-2}$)     &   $<16.7 $           \\
         &        log(\oiii) \ ($\rm cm^{-2}$)  &  $<15.8  $         \\
          &   log(\oiv) \  ($\rm cm^{-2}$)       &  $ <16.4  $        \\
\hline
 \amol     &        $\rm N_{dust} (\rm 10^{16} \ cm^{-2}) $    &     $8.13^{+0.92}_{-1.07}$   \\
\hline
    \end{tabular}
    \caption{Best fit parameters for the oxygen spectral region. The values are a result of the simultaneous fit of the FUV and X-ray spectral lines using \spex models and emcee. \NH is the neutral hydrogen column density in the line of sight; $\rm Z_{v}$ is the velocity shift of the absorber; $\rm log(norm_{bb})$ is the logarithm of the black body normalisation; $\rm N_{dust}$ is the total column density of dust; $b$ refers to the line velocity broadening; \oii, \oiii and \oiv correspond ionic column densities from the \slab model; $O_{abund}$ is the total O abundance scale factor relative to the abundance table of \cite{Lodders2009}; and \oi is the implied column density of this ion from the \hot model.} 
    \label{tab:oxygen}
\end{table}

\section{Discussion}
\label{discussion}

The \HSTCOS spectrum reveals at least two discrete absorbers with distinct velocity components. As shown in Figure \ref{fig:cosdata}, the interstellar absorption lines from the \feii \ ion indicate two blue-shifted components, one with velocity shift of $-37\pm4 $ km/s and one with $-96\pm5$ km/s. Similarly, \sii \ shows two components, one at $-20\pm2 $ km/s and one at $-84\pm2$ km/s away from the rest-frame velocity. 
We used the kinematic distance calculator of \citet{Reid2014}\footnote{\url{http://bessel.vlbi-astrometry.org/revised_kd_2014}} to estimate the distance to the \feii and \sii absorbers.  
Our analysis revealed that the components corresponding to \feii are situated at distances of approximately $2.38\pm0.35$ kpc and $5.01\pm0.3$ kpc, while for \sii, we found distances of $1.44\pm0.45$ kpc and $4.5\pm0.3$ kpc, respectively.
The velocity shifts and cloud distances of \sii \ and \feii components show some inconsistency within the margins of error. However, given their close proximity and the similarity in their ionization potentials, it's plausible that they arise from similar locations. If this is the case, then those two systems are likely $\approx 1-3$~kpc and $4-5.5$~kpc away.
For the singly ionised form of carbon \cii, we were able to constrain a velocity shift of the first spectral component to $-19\pm2$ km/s, likely associated with the nearer \sii system. The second component is too weak to obtain a good fit when the line-of-sight velocity is left as a free parameter, so this value was frozen to -95 km/s in order to obtain a column density measurement.

Our discussion is organised into several parts. First, we review existing literature on standard abundance tables for the elements we're studying, and we use neon as a reference point for comparison. Next, we introduce a \cloudy grid that helps us compare our findings from both X-ray and FUV data regarding the column density ratios of different ions. Then, we present individual discussions for each element. Finally, we highlight discrepancies in the atomic data used for iron K-shell absorption in the literature.

\subsection{Elemental abundances across the literature and \cloudy}
\label{abundances_cloudy}

\begin{figure*} [ht]
\begin{centering}
\includegraphics[width=0.55\textwidth]{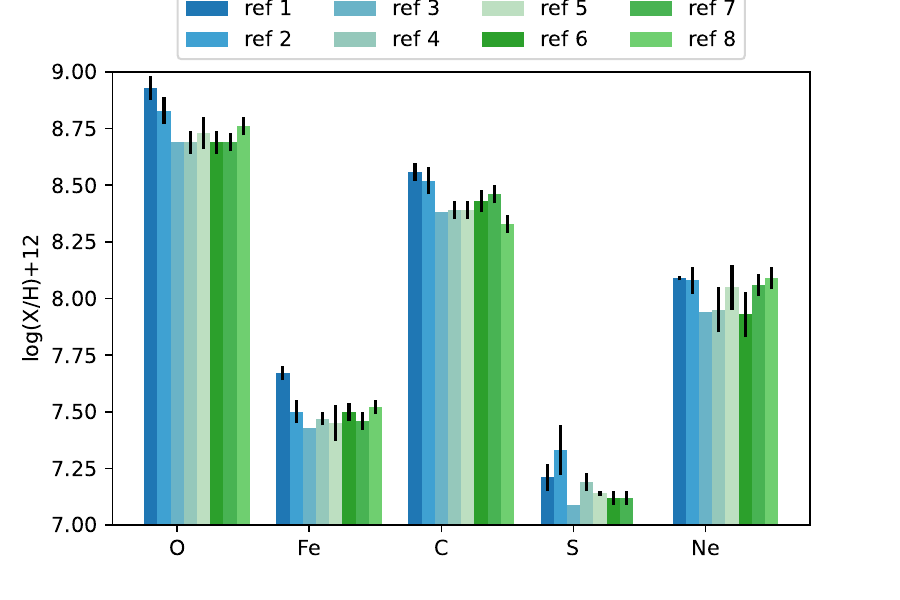}
\caption{Comparison of the different abundance tables for every element, where ref 1:\cite{Anders1989}, ref 2:\cite{Grevesse1998}, ref 3:\cite{Wilms}, ref 4:\cite{Lodders2003}, ref 5 :\cite{Lodders2009}, ref 6:\cite{Asplund2009}, ref 7:\cite{Asplund2021}, ref 8:\cite{Nieva2012}.} 
\label{fig:abundances}
\end{centering}
\end{figure*}

In Table \ref{tab:abundances} we report the abundance of O, Fe, C, S, and Ne among the different values found in the literature, in standard units of $\rm log(X/H)+12$, where X/H represents the abundance of each element with respect to hydrogen. 
\cite{Anders1989} present abundance tables for both meteoric and solar photosphere data. For our comparisons we assume the photospheric values, although the two sets are generally consistent with each other, except a few elements. For Fe the solar value is $7.67\pm0.03$, while the meteoric is $7.51\pm0.01$. Similarly, the tabulated values from \cite{Grevesse1998}, \cite{Lodders2003}, \cite{Asplund2009} and \cite{Asplund2021} refer to the solar photospheric values. 
\cite{Wilms} present a model for the absorption of X-rays in the ISM. The selected values come from the adopted abundance of the ISM based on \cite{Snow1996}, \cite{cardelli1996} and \cite{Meyer1998}. Lastly, we also include in the comparison B-type star elemental abundances from \cite{Nieva2012}.
In this study, our spectral models are based on the proto-Solar abundances as provided by \cite{Lodders2009}, which serve as the default abundance set in \spex. In Figure \ref{fig:abundances}, we present a comparative analysis of the elemental abundances listed in Table \ref{tab:abundances} for the elements under investigation in our study. 
Variations in the reported values make it important to note that the choice of the reference abundance table can have an impact on the resulting measured abundances. 

Because it is a noble gas, neon will not deplete into dust grains. Thus it can serve as a suitable reference for comparing the observed abundance of elements, providing an alternative to hydrogen, which does not provide any spectral features in the X-ray band. The last four columns of Table \ref{tab:abundances} display the calculated X/Ne ratio for each of the literature abundance tables. The Ne abundance is determined through the fit of the Ne K-edge in the X-ray spectrum from \cite{Psaradaki2022}, and it represents the summed abundance of \nei, \neii and \neiii, which is $\rm 2.1\times 10^{17} \ cm^{-2}$ . 
In the final row of Table \ref{tab:abundances}, we present our computed value for log(X/Ne). 
In Figure \ref{fig:abundances}, we visualise the results of Table \ref{tab:abundances}, showing the deviations of the literature standard abundance tables for Fe, O and S compared to neon, with the values of this work derived from FUV and X-ray observations.
We will revisit this comparison of elemental abundances with neon in the following sections, where we will explore individual discussions of the elements under study. 

We further use the spectral synthesis code \cloudy  \cite[version 2017,][]{ferland17}, and ran a grid of models over a wide range of ionization parameter (ionizing photon density) and metallicity values for a neutral Hydrogen column density of 2$\times$10$^{21}$ cm$^{-2}$. The ionization parameter $U$ is defined as the ratio between the number densities of ionizing photons and hydrogen ($U \equiv n_{\gamma} / n_{\rm H}$), and we allow this parameter to vary from $\log{U} = -4$ to 0 by 0.25 dex. We also vary $\log{Z/Z_{\odot}} = -2$ to +0.5 in steps of 0.1 dex. 
We adopt a photoionizing spectrum from the Milky Way that includes a contribution from the extragalactic UV background \citep{fox05} and assume thermal and ionization equilibrium for a plane-parallel slab geometry with a uniform density. 
We use the ``grains ISM" command to specify grains with a size distribution and abundance consistent with those in the Milky Way and additionally employ the ``metals deplete" command in our models  to deplete elements that are included in grains, e.g. Fe,  according to the work of \citet{Jenkins2009}. 
We utilize the \cloudy grids to compare between the ratios of the ions investigated in this study and the corresponding values predicted by \cloudy.
We found that the relative abundances of each gas-phase ion were relatively insensitive to the magnitude of ionization parameter $U$ for the photoionising spectrum used in this run. We use logU=-5 as our fiducial value in the discussion below.

First, we compare our observed ion abundances for oxygen with those predicted by \cloudy, in scenarios where elements are not depleted into dust. This approach allows us to assess the predicted ion quantities in the case of excluding the influence of the dust phase, thereby discerning any disparities. Using \cloudy, we have determined that \oii/\oi=$\rm 4.62 \times 10^{-4}$ and \oiii/\oi=$\rm 1.2 \times 10^{-7}$. Neutral oxygen gas is thereby expected to significantly dominate the neutral ISM abundance when compared to \oii and \oiii, with \oiii being the least abundant among these ions. However, \cloudy functions as a photoionized model, diverging from the primarily collisionally ionized models used in the \hot model of \spex, and one should exercise caution when comparing to the absolute value of the ionic ratios. 
Through simultaneous fitting of both X-ray and UV data, we established upper limits for the ionic column densities of \oii, \oiii, and \oiv ions (Table \ref{tab:oxygen}). With the exception of \oiii, these upper limits are consistent with the derived values of \cite{Gatuzz2018} for the same source.

\begin{table}
\caption{Literature standard abundances for the elements in this study. The standard abundances are expressed in logarithmic units, with hydrogen by definition 12.0.}  
\label{tab:abundances} 
\begin{center}
\resizebox{\textwidth}{!}{\begin{tabular}{ccccccccc}
\hline\hline             
Reference                                       &   $\rm log(O/H)+12$      & $\rm log(Fe/H)+12$     &  $\rm log(C/H)+12$ & $\rm log(S/H)+12$   &  $\rm log(Ne/H)+12$ & log(O/Ne) & log(Fe/Ne) & log(S/Ne) \\
\hline   
\cite{Anders1989}						    	&      $8.93\pm0.035$     &      $7.67\pm0.03$       & $8.56\pm0.04$       & $7.21\pm0.06$   &  $\rm 8.09\pm0.10$  & $0.84\pm0.10 $&  $-0.42\pm0.10 $& $-0.88\pm0.01$    \\
\cite{Grevesse1998}                             &    $8.83\pm0.06$         &  $7.5\pm0.05$              &   $8.52\pm0.06$    &  $7.33\pm0.11$  &  $8.08\pm0.06$  & $0.75\pm0.08 $& $-0.58\pm0.08 $& $-0.75\pm0.11 $\\
\cite{Wilms}\footnote{Note that these values come from the adopted abundance of the ISM based on \cite{Snow1996}, \cite{cardelli1996} and \cite{Meyer1998}. }     & 8.69     &     7.43     &   8.38  &   7.09     &    7.94       & 0.75  & -0.51 & -0.85 \\
\cite{Lodders2003}                               &    $8.69\pm0.05$      &     $7.47\pm0.03$          &    $8.39\pm0.04$    &  $7.19\pm0.04$   & $7.95\pm0.10$  & $0.74\pm0.11 $ &$-0.48\pm0.10$ &$-0.76\pm0.10$ \\
\cite{Lodders2009}\footnote{Used in this study, and default set of abundances in \spex.}                               &    $8.73\pm0.07$        &   $7.45\pm0.08$     &  $8.39\pm0.04$      & $7.14\pm0.01$  & $8.05\pm0.10$ & $0.68\pm 0.12$ & $-0.6\pm0.13$ &  $-0.91\pm0.10$    \\
\cite{Asplund2009}\footnote{We refer to the photospheric values.}       &       $8.69 \pm 0.05 $   &   $7.5\pm0.04$         &    $8.43\pm0.05$   &    $7.12\pm 0.03$  & $7.93\pm0.10$ & $0.76\pm0.11 $&$-0.43\pm0.1$ & $-0.81\pm0.10$\\
\cite{Nieva2012}\footnote{Abundances derived from B-type stars.}       &       $8.76 \pm 0.04 $   &   $7.52\pm0.03$         &    $8.33\pm0.04$   &    -  & $8.09\pm0.05$ & $0.67\pm0.06$ & $-0.57\pm0.06$ & - \\
\cite{Asplund2021}                              &     $8.69 \pm 0.04 $     &  $7.46\pm0.04$           &    $8.46\pm0.04$      &     $7.12\pm0.03$    & $8.06\pm0.05$   & $0.63\pm0.10$ & $-0.60\pm0.06$  &$-0.94\pm0.06$ \\
\hline
This study\footnote{The log(O/Ne) ratio is the sum of the predicted \oi abundance that we obtain from a simultaneous X-ray and FUV fit, and the solid phase oxygen. For the log(Fe/Ne) ratio, we used the combined contribution of solid iron and atomic \feii. The log(S/Ne) ratio comes from the \sii value only resulting from the FUV fit. The Ne abundance is determined through the fit of the Ne K-edge in the X-ray spectrum from \cite{Psaradaki2022}, and in particular it represents the summed abundance of \nei, \neii and \neiii.} &                     &                     &                     &                   &               &  $0.81\pm0.03$               &$-0.41\pm0.11$  &   $-1.35^{0.04}_{0.05}$             \\
\hline
\end{tabular}
}
\end{center}
\end{table}

\begin{figure*} [ht]
\begin{centering}
\includegraphics[width=1\textwidth]{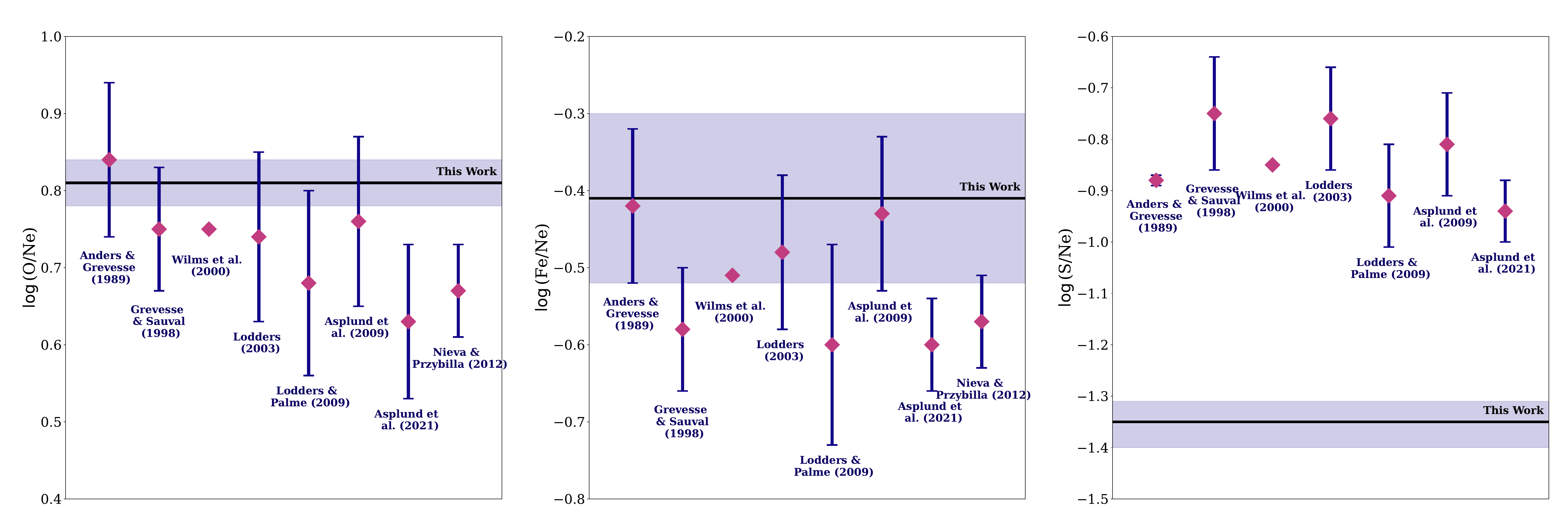}
\caption{Comparison of the literature standard abundance tables of Fe, O and S with the values of this work derived from FUV and X-ray observations, presented in Table \ref{tab:abundances}. The log(O/Ne) ratio is the sum of the predicted \oi abundance that we obtain from a simultaneous X-ray and FUV fit, and the solid phase oxygen. For the log(Fe/Ne) ratio, we used the combined contribution of solid iron and atomic \feii. Please note that the log(S/Ne) ratio comes from the \sii value only resulting from the FUV fit. The Ne abundance is determined through the fit of the Ne K-edge in the X-ray spectrum from \cite{Psaradaki2022}, and in particular it represents the summed abundance of \nei, \neii and \neiii.     } 
\label{fig:abundances_Ne}
\end{centering}
\end{figure*}

\subsection{The abundance and depletion of Fe}

Understanding the exact reservoirs of iron in the diffuse ISM is still an open question. 
Our X-ray fits yield a column density of solid-phase iron of $\rm 3.9\times10^{16} \ cm^{-2}$. The atomic component is in \feii\ and contributes to the total column density through two distinct absorption systems (Table \ref{spexparams}), amounting to $\rm 2\times 10^{15} \ cm^{-2}$. 
Our \cloudy simulations, detailed in Section \ref{abundances_cloudy}, indicate that the \fei/\feii ratio should be approximately $\rm 2.5\times 10^{-3}$, while the \feiii/\feii ratio is on the order of $\rm 10^{-6}$. These findings suggest that \feii is the dominant form of gas-phase iron in the neutral ISM, with \fei and \feiii concentrations being negligible. 
The abundance of \feii is still relatively small compared to the abundance of iron in solid form, accounting for merely 4\% of the total iron content. The remaining 96\% resides in solid-state structures, in the form of amorphous pyroxene ($\rm Mg0.75Fe0.25SiO_{3}$), and metallic iron. This dust grain mineralogy was previously found in \cite{Psaradaki2022}, and it is assumed in this study. However, the dust column density of each compound is free, and it is found to be $\rm <6.1\times10^{17} \ cm^{-2}$ and $\rm <1.2\times10^{16}\ cm^{-2}$ for the amorphous pyroxene and metallic iron respectively. Additionally, we examined the scenario where Fe exists principally in its metallic form, as detailed in \cite{Westphal} for the observation of Cygnus X-1. However, when applying this model to the case study of Cygnus X-2, we observed a less optimal fit. The investigation of metallic iron as a compound needs further investigation, primarily due to uncertainties in the energy calibration across various measurements of this compound in the literature (\citealt{Psaradaki2021}, Costantini et al. in prep., Corrales at al. submitted).

The selection of the MRN dust size distribution could potentially have an effect in the calculation of the dust column density. In particular, the phenomenon of self-shielding may play a role in diminishing the overall iron column available for photoelectric absorption (\citealt{Wilms}).  In this case, strong absorption prevents X-rays from penetrating the inner portions of the dust grain, and a smaller fraction of the total metal column contributes to the absorption edge (\citealt{Corrales2016}). Our study incorporates self-shielding, and we have considered the extinction (scattering + absorption) cross section in our spectral modeling. However, this effect is anticipated to be particularly noticeable in regions of the interstellar medium (ISM) containing large grains, approaching the upper limit of the dust size distribution employed in our investigation. Consequently, it is plausible that some of the depleted iron is located within populations of large grains, specifically those exceeding 0.25 $\mu$ m in size. Other size distributions beyond the MRN, such as those employed in \citet{Zubko2004} and \citet{Weingartner2001}, will be examined in a future study.

We extend our calculations to determine the combined abundance of iron in both gaseous and dust components, comparing it to established standard abundance values found in the literature. The third column in Table \ref{tab:abundances} presents a comparison of the iron abundance figures, denoted in units of $\rm log(Fe/H)+12$. Taking into account the associated uncertainties, our iron abundance estimation from our best fit is $7.38\pm0.33$, which takes into consideration iron in dust (comprising silicates and metallic iron) and \feii. This result is in agreement with the standard abundance tables.
Moreover, in Table \ref{tab:abundances} and Figure \ref{fig:abundances_Ne} left panel, we show a comparison of the iron abundance tables compared to neon. 
Overall there is consistent behaviour between our calculated $\rm log(Fe/Ne)$ value, derived from spectral fitting of X-ray and UV data, and the standard abundance tables reported in the literature, with the exception of the most recent work of \citet{Asplund2021}. 

\subsection{Where is sulfur?}

In the COS spectrum of Cygnus X-2, we have identified singly ionized atomic sulfur lines (\sii) at wavelengths 1250.58 \AA\ and 1253.81 \AA. These spectral lines are associated with two distinct velocity clouds. The first cloud exhibits a column density of approximately $(7.4\pm0.5) \cdot 10^{15} \ \rm cm^{-2}$
and the second cloud shows a weaker transition with a column density of approximately $(1.9\pm0.1) \cdot 10^{15} \ \rm cm^{-2}$. 
When examining Table \ref{tab:abundances} and the right panel of Figure \ref{fig:abundances_Ne}, we observe that the SII/Ne ratio derived from our analysis is under-abundant compared to the total S/Ne ratio calculated from the literature. This suggests that there should be another reservoir of sulfur other than \sii in the diffuse sight-line of Cygnus X-2. 

\cloudy modeling further predicts that the \si/\sii ratio is $\rm 2.8\times 10^{-4}$, implying that the remaining sulfur is not in the neutral gas-phase. Similarly, the \sii/\siii ratio is $\rm 7.1\times 10^{-4}$. These predictions imply that the remaining sulfur is not expected to be in the form of \si or \siii; instead, it could be bound within molecules or dust particles. However, it is essential to exercise caution when utilizing the \cloudy ratios in this context. \cloudy operates as a photoionized model, presenting a contrast to the predominantly collisionally ionized models employed in the \hot model of \spex, and widely adopted in previous studies.
Moreover, in \cite{Gatuzz2024} the S K-edge has been examined using high-resolution \chandra/HETGS spectra of 36 low-mass X-ray binaries. In the case of Cygnus X-2 their ionic column density estimates appear to disagree with the \cloudy predictions. However, only upper limits of the ionis column densities were able to be provided.

Sulfur in dust species can take on various forms, including FeS, $\rm FeS_{2}$, or even exist within GEMS (\citealt{Bradley1994}), where the FeS particles would be more concentrated on the surface of the glassy silicate. However, studies have demonstrated that GEMS are less favored as a plausible component of interstellar dust (\citealt{Keller}, \citealt{Keller2011}, \citealt{Westphal}. In \cite{Psaradaki2022} we used newly computed dust extinction models of astrophysical dust analogues for the Fe L-edges including FeS or $\rm FeS_{2}$. However, strong evidence for these species was not found in those works. 
It has been discussed in the literature that sulfur does not appear to change depletion in the diffuse ISM, suggesting that it does not easily get incorporated into dust (\citealt{Sembach1996}). However, in molecular clouds, sulfur can be included in aggregates such as for example $\rm H_{2}S$ or $\rm SO_{2}$ (\citealt{Duley1980}). Inclusion into simple atomic sulfur or sulfur ices have been proposed to solve the missing-sulfur problem in dense molecular clouds (\citealt{Vidal2017}). 
We examined the three-dimensional maps of interstellar dust reddening, which are based on Pan-STARRS 1 and 2MASS photometry, and Gaia parallaxes\footnote{\url{http://argonaut.skymaps.info/}} (\citealt{Green2019} and references therein). These maps trace the dust reddening both as a function of angular position on the sky and distance. Using these maps, we did not find any steep jump in the line-of-sight reddening. This could suggest that the line of sight towards Cygnus X-2 is rather diffuse, and does not cross a dense molecular cloud. Thus the nature of the missing sulfur in the Cygnus X-2 sight line, as determined in this study, is still a mystery.

A comprehensive understanding of sulfur depletion within dust particles can be achieved through the examination of the sulfur K-edge at 2.48~keV in X-ray spectra. Unfortunately, the column density towards Cygnus X-2 falls short in providing the necessary optical depth to study the photoabsorption edge of sulfur. Moreover, the current X-ray instruments utilized in this study lack sufficient energy resolution at this critical energy range. The recently launched X-ray Imaging and Spectroscopy Mission (XRISM) will enable us to study directly the photoabsorption edge of sulfur \cite[e.g.][]{Costantini2019}, and determine the dust inclusion of this element. 
\subsection{The carbon abundance}

The gas-phase carbon in the neutral ISM could be primarily
in the form of singly ionised species, \cii, because the \ci ionization energy is lower than that of $\rm H_{I}$, and the \cii ionization energy is above that of $\rm H_{I}$. Using \cloudy we find indeed that the \ci/\cii ratio is $\rm 3.1\times 10^{-3}$.
Surprisingly, we measure comparable column densities of \ci and \cii from the FUV spectral fit. We find that the total column density of the \cii absorbers in the line of sight is $\rm (4.7\pm0.3) \times 10^{14} \ cm^{-2}$, while for \ci we find $\rm 3.6^{+1.1}_{-0.6} \times 10^{14} \ cm^{-2}$.

We compare these values with carbon-related results available in the existing literature. \cite{Gatuzz2018} studied the C-K edge using high-resolution Chandra spectra of four
novae during their super-soft-source state. They have found column densities of \cii in the range of $\rm (1.8-3.5)\times10^{17} \ cm^{-2}$, which is inconsistent with our values in Table \ref{spexparams}. Moreover, in the study by \cite{Sofia_CII}, the \cii \ (2325\AA) equivalent width was measured in an absorption system directed toward the diffuse sightline of the $\tau$ Canis Majoris star. The results indicated a column density of $\rm (7.57 \pm 2.52) \times 10^{16} \ cm^{-2}$ for this system (and $10^{6}$ \cii / \hi =$135\pm46$). This finding was later complemented by \cite{Sofia2009}, who investigated various sightlines in the interstellar medium with known hydrogen abundances, utilizing HST/STIS data. Through the modeling of the strongest lines, they found \cii column densities ranging from $\rm (1.97-6.19)\times10^{17} \ cm^{-2}$ across different lines of sight.
In addition, \cite{Cardelli1993} detected \cii in diffuse clouds toward $\zeta$ Oph using the Goddard High Resolution Spectrograph, reporting a column density of $1.8 \times 10^{17} \ \mathrm{cm^{-2}}$.
Similarly, \cite{Cardelli1991b} examined observations of ultraviolet interstellar absorption lines of dominant ion stages arising in the diffuse clouds in the direction of $\xi$ Persei, and focused on the same \cii line, reporting a column density of $\rm 5\times 10^{17} \ cm^{-2}$. Collectively, these studies offer insights into the variations of \cii column densities across different interstellar absorption systems.
The column density of \cii absorption observed in this study is significantly lower compared to the values reported above. 
One explanation could be that we are most likely probing \cii that has been ionized by the ambient interstellar radiation field at the edges of the \sii bearing cloud, rather than a large photoionization region, as is expected around the massive O-type stars examined in the above works.

The \ci in our study shows a different trend compared to the other ions studied here. The first and most dominant component shows a velocity shift of -25~km/s, between that of the nearer \sii and \feii absorbing region. Perhaps this stronger \ci absorption could be arising from denser regions of the ISM that are shielded from the interstellar radiation field. 
The second, weaker component is red-shifted compared to the rest-frame velocity. The source of this red-shifted \ci is unknown. 
\cite{Jenkins2011} studied the UV spectra of 89 stars using HST data. Based on the integrated \ci absorption across all velocities, they determined that the column density of the \ci absorbers falls within the range of approximately $\rm 2.4\times 10^{13}\ cm^{-2}$ to $\rm 5.7\times 10^{14}\ cm^{-2}$ (lower limit). Our findings align with these results. 

There are still uncertainties around the abundance and depletion of carbon within dust grains. Although carbon is a substantial element in grains, our understanding of the mechanisms through which dust grains incorporate carbon remains rather incomplete (\citealt{Jenkins2009}). This topic continues to be an active area of study. Future advancements, including upcoming X-ray missions and innovative concepts like Arcus (\citealt{Arcus}), hold the potential to carry out in-depth spectroscopic analysis around the C K-edge, and in particular the features of dust as well as \ci (\citealt{Costantini2019}). 

\subsection{Discrepancies in the available Fe X-ray atomic data}

In high-resolution X-ray spectroscopy the choice of atomic data plays a crucial role in the analysis of the data and the interpretation of the results. In \citet{Psaradaki2020} we discussed the discrepancy between the atomic databases of \spex and \xstar for the oxygen ions, from \oi to \oiv. We found out that the calibration of the energy scale of the different models can differ, and this can have an effect on the results, especially with future X-ray telescopes. 

\begin{figure}
\centering
\includegraphics[width=0.4\linewidth]{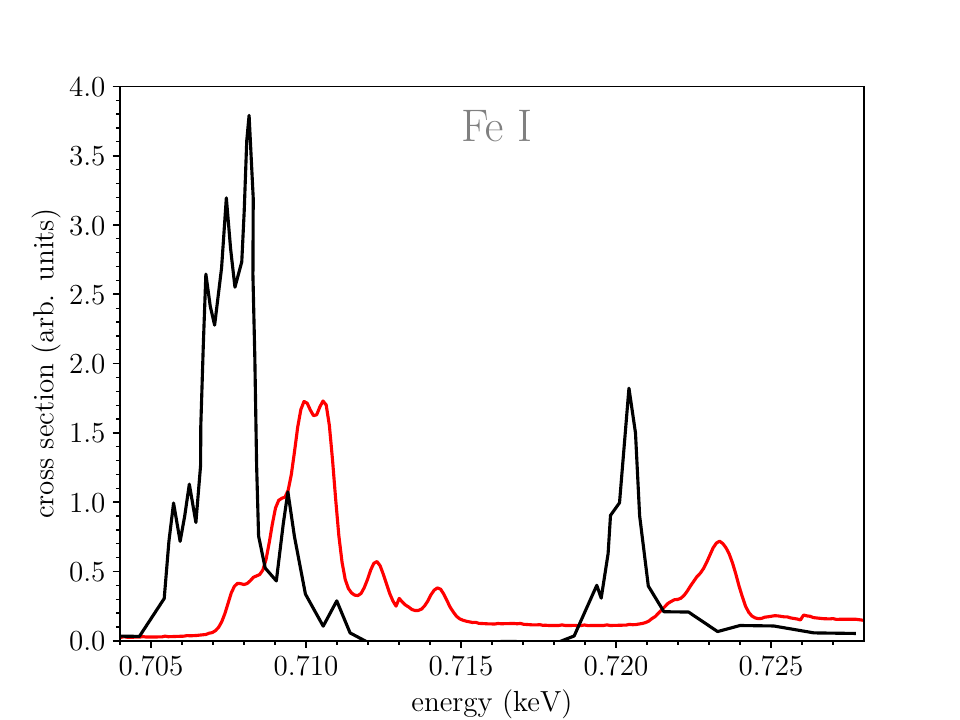}
    \includegraphics[width=0.4\linewidth]{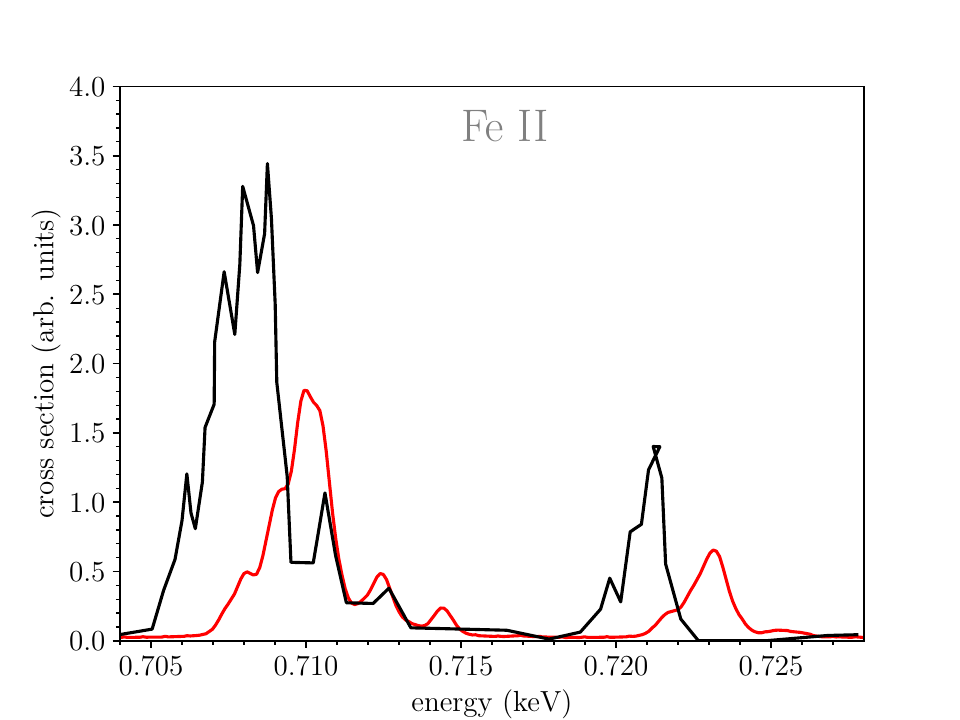}
    \includegraphics[width=0.4\linewidth]{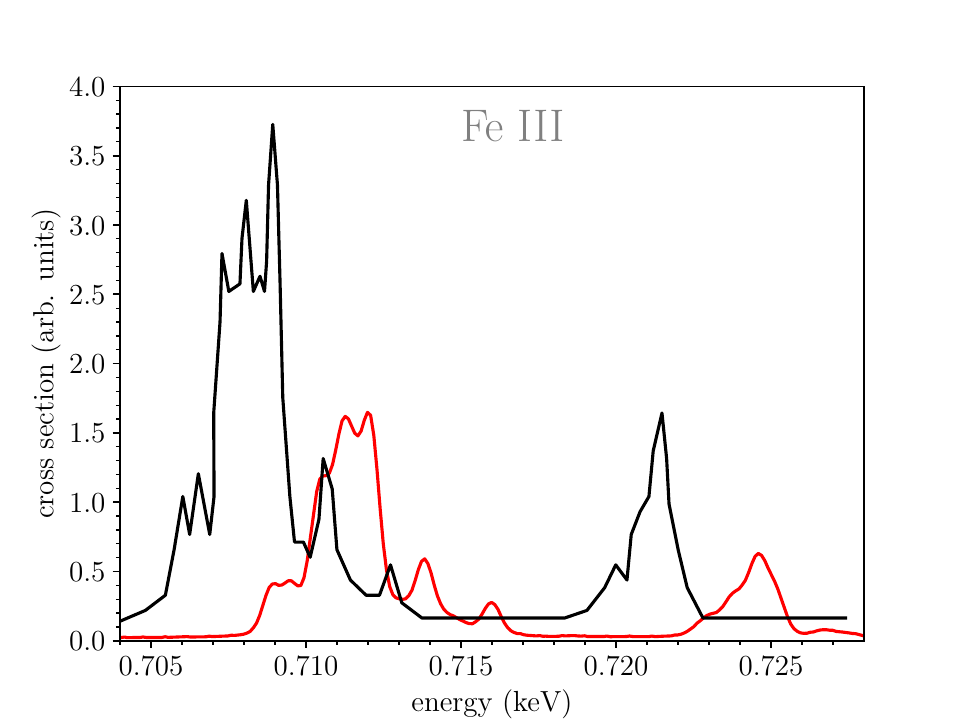}
    \includegraphics[width=0.4\linewidth]{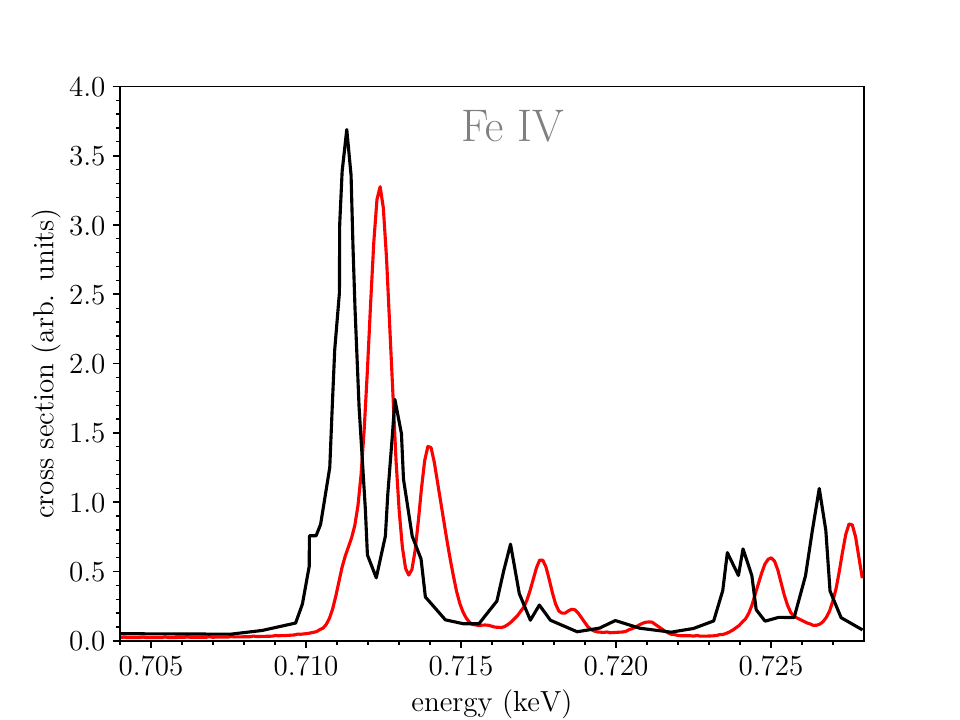}
\caption{Comparison between the X-ray atomic database  of \spex (red) and \citet{Schippers2021} (black) for \fei, \feii, \feiii, \feiv. The \fei, \feii and \feiv data in \citet{Schippers2021}, have been taken from \citet{Richter2004}, \citet{Schippers2017}, \citet{Beerwerth2019}, respectively.}
\label{fig:atomic}
\end{figure}

In this section we examine the discrepancy in the atomic database for iron. We compare the atomic data implemented in \spex \ with the available data of \fei \ - \feiv ions presented in the recent work of \citet{Schippers2021}. The \fei, \feii and \feiv data in \citet{Schippers2021}, have been taken from \citet{Richter2004}, \citet{Schippers2017}, \citet{Beerwerth2019}, respectively. In figure \ref{fig:atomic} we compare the databases. From the plots it is evident that there is a shift of about 2.7 eV between the \spex atomic data and the models presented in \citet{Schippers2021}, which is detectable with the energy resolution of the \chandra \ HETGS instrument. 

We tested how the discrepancy between the atomic databases can affect the X-ray spectral fitting of Cygnus X-2. We shifted the absolute energy of the \fei - \feiv \ transitions in \spex  according to the energy calibration reported in \citet{Schippers2021}. We kept the same model described in Section \ref{xray_fit}, and repeated the fit. Around the Fe L-edges the X-ray absorption is dominated by the dust, while iron in atomic form is too weak to constrain the X-ray fits. We therefore achieved similar results. These discrepancies however will be more evident with future X-ray instruments, such as with the spectral resolution capabilities demonstrated by the grating spectrometers of Arcus (\citealt{Arcus}). 

\section{Conclusions}
\label{conclusions}

In this work we combined high-resolution X-ray and FUV spectroscopic data from \chandra, \xmm and the Hubble Space Telescope. Our primary goal was to gain insights into the abundance and depletion patterns of oxygen, iron, sulfur, and carbon. To achieve this, we developed a novel analysis pipeline that involves a combined fitting of UV and X-ray datasets. This approach incorporates the consideration of the Line Spread Function (LSF) of \HSTCOS for more accurate results. Our main conclusions can be summarized as follows:

\begin{itemize}

\item The Hubble Space Telescope's COS spectrum has unveiled intriguing insights into the line of sight toward Cygnus X-2. 
Our investigation of various ions, including \feii, \sii, and \cii, has led to the identification of at least two distinct absorption systems, each exhibiting blue-shifted velocity components. From the kinematics of the known Milky Way, these line-of-sight velocities correspond to ISM regions that are 1-3 and 4-5.5 kpc away. Neutral carbon presents an anomaly. The strongest absorption line has a blue-shifted velocity consistent with the nearer absorption system. However, we also observe a red-shifted velocity component, the source of which is unknown.

\item \cloudy simulations suggest that the majority of gaseous-phase interstellar iron should predominantly exist in the form of \feii, with $\lesssim 10^{-3}$ of the gas-phase iron contributions coming from \fei and \feiii. Moreover, we find that our derived iron abundance, accounting for iron present in \feii and dust, which comprises silicates and metallic iron, is consistent with the solar values from the literature listed in Table \ref{tab:abundances}.

\item \cloudy simulations suggest that \sii is expected to be the dominant gas-phase ion of sulfur, rather than \si or \siii. However, the abundance of \sii directly measured from the \HSTCOS data is much lower than expected from standard abundance arguments. It is apparent that an additional repository for sulfur is needed, possibly in the form of dust. Intriguingly, most X-ray analyses do not find strong signatures of FeS compounds \citep[][Corrales et al., submitted]{Westphal, Psaradaki2022}, prompting the exploration of alternative compounds. 
The X-ray Imaging Spectroscopy Mission \citep[XRISM, launched September 2023,][]{Tashiro2020} has the collecting area and energy resolution to potentially resolve this issue. In particular, investigating the sulfur K and iron K edges simultaneously could unlock this mystery.

\item The X-ray atomic databases employed in high-resolution X-ray spectroscopy, particularly in the vicinity of the Fe L-edges, may be a source of additional uncertainty. Notably, deviations in energy scale have been observed, with discrepancies of up to 2.7 eV. 
These disparities will become even more pronounced in the context of future X-ray instruments with enhanced spectral resolution within the soft X-ray range, such as the Arcus concept mission (\citealt{Arcus}).
\end{itemize}

\noindent In conclusion, our investigation on the depletion and abundances of Fe, O, S and C demonstrates the potential of combining X-ray and FUV data. This is a powerful way to determine the abundance of these elements in atomic form, and then estimating their presence in dust species through high-resolution X-ray spectra. It is therefore encouraged to extend this study to more sightlines along the Galactic plane.

\begin{acknowledgements}

We thank the referee for the suggestions that helped to improve this paper. The authors would like to thank J. de Plaa for help with \spex. 
This research has been supported by NASA’s Astrophysics Data Analysis Program, grant number 80NSSC20K0883, under the ROSES program NNH18ZDA001N. Support for this work was also provided by NASA through the Smithsonian
Astrophysical Observatory (SAO) contract SV3-73016 to MIT for Support
of the Chandra X-Ray Center (CXC) and Science Instruments. CXC is
operated by SAO for and on behalf of NASA under contract NAS8-03060
\end{acknowledgements}

Some of the data presented in this article were obtained from the Mikulski Archive for Space Telescopes (MAST) at the Space Telescope Science Institute. The specific observations analyzed can be accessed via \dataset[DOI: 10.17909/vdjz-xn16]{https://doi.org/10.17909/vdjz-xn16}. 
This paper also employs a list of Chandra datasets, obtained by the Chandra X-ray Observatory, contained in~\dataset[DOI: 10.25574/cdc.202]{https://doi.org/10.25574/cdc.202}.

\bibliography{reference}
\bibliographystyle{aasjournal}

\end{document}